\documentclass[12pt,preprint]{aastex}
\usepackage{amsmath}

\newcommand{\et}{et al.}
\newcommand{\kms}{km s$^{-1}$}
\newcommand{\ha}{H$\alpha$}

\newcommand{\solar}{\ifmmode_{\sun}\;\else$_{\sun}\;$\fi}

\newcommand{\HI}{\ion{H}{1}}
\newcommand{\HII}{\ion{H}{2}}

\newcommand{\mhilb}{M$_{HI}$/L$_B$}

\newcommand{\vii}{VII Zw 403}

\newcommand{\hres}{B$+$C$+$D$_{-1}$}
\newcommand{\mres}{B$+$C$+$D$_0$}
\newcommand{\lres}{C$+$D$_1$}
\newcommand{\bcdone}{B$+$C$+$D$_1$}

\newcommand{\msun}{$M_{\sun}$}
\newcommand{\acm}{cm$^{-2}$}

\begin{document}

\slugcomment{To Appear in the Astronomical Journal}

\title{VII Zw 403: \HI\ Structure in a Blue Compact Dwarf Galaxy}

\author{Caroline E.\ Simpson}
\affil{Department of Physics, Florida International University, Miami,
  Florida 33199 USA} 

\email{simpsonc@galaxy.fiu.edu}

\author{Deidre A.\ Hunter}
\affil{Lowell Observatory, 1400 West Mars Hill Road, Flagstaff, Arizona
   86001 USA}

\author{Tyler E.\ Nordgren}
\affil{Department of Physics, University of Redlands, 1200 East Colton
  Avenue, Redlands, CA 92373 USA}

\author{Elias Brinks}
\affil{Centre for Astrophysics Research, University of Hertfordshire,
  Hatfield AL 10 9AB UK} 

\author{Bruce G.\ Elmegreen}
\affil{IBM T.\ J.\ Watson Research Center, PO Box 218, Yorktown Heights,
  New York 10598 USA} 

\author{Trisha Ashley}
\affil{Department of Physics, Florida International University, Miami, Florida 33199 USA}

\author{Roger Lynds}
\affil{Kitt Peak National Observatory, NOAO, Box 26732, Tucson, AZ 85726 USA}

\author{Vince J.\ McIntyre}
\affil{Australia Telescope National Facility, CSIRO, P.O. Box 76,
  Epping, NSW 1710, Australia} 

\author{Earl J.\ O'Neil}
\affil{Kitt Peak National Observatory, NOAO, Box 26732, Tucson, AZ 85726 USA}

\author{G\"oran \"Ostlin}
\affil{Stockholm Observatory, Department of Astronomy, AlbaNova,
  SE-10691, Stockholm, Sweden} 

\author{David J.\ Westpfahl}
\affil{Department of Physics, New Mexico Institute of Mining and
  Technology, Socorro, NM 87801 USA} 

\author{Eric M.\ Wilcots}
\affil{Washburn Observatory, University of Wisconsin, 475 N. Charter
  St., Madison, Wisconsin 53706 USA} 

\begin{abstract}

We present optical ($UBVJ$), ultraviolet (FUV, NUV), and high resolution
atomic hydrogen (\HI) observations of the nearby blue compact dwarf
(BCD), VII Zw 403. We find that VII Zw 403 has a relatively high \HI\
mass-to-light ratio for a BCD.  The rotation velocity is nominally
$10-15$ \kms, but rises to $\sim 20$ \kms\ after correction for the
$\sim 8-10$ \kms\ random motions present in the gas. The velocity field
is complex; including a variation in the position angle of the major
axis going from the NE to the SW parts of the galaxy. Our high
resolution \HI\ maps reveal structure in the central gas, including a
large, low-density \HI\ depression or hole between the southern and
northern halves of the galaxy, coincident with an unresolved x-ray
source.

Although interactions have been proposed as the triggering mechanism
for the vigorous star formation occurring in BCDs, VII Zw 403 does not
seem to have been tidally triggered by an external interaction, as we
have found no nearby possible perturbers. It also doesn't appear to
fall in the set of galaxies that exhibit a strong central mass density
concentration, as its optical scale length is large in comparison to
similar systems. However, there are some features that are compatible
with an accretion event: optical/\HI\ axis misalignment, a change in
position angle of the kinematic axis, and a complex velocity field.

\end{abstract}

\keywords{galaxies: irregular --- galaxies: star formation ---
galaxies: individual: VII Zw 403 ---
galaxies: ISM}

\section{Introduction}

VII Zw 403 is classified as a blue compact dwarf (BCD) galaxy.
Compact galaxies were first identified by \citet{zwicky64} as high
surface brightness objects that were just distinguishable from stars on
the Palomar Sky Survey.  This original classification included galaxies
with a wide range of colors and total luminosities. Later,
\citet{tm81} presented a study of 115 blue compact dwarf galaxies,
which they defined as low luminosity ($M_B \leq -18$) galaxies with a 
compact optical size ($\lesssim 1$ kpc in diameter), that have optical
spectra with ``strong sharp narrow emission lines superposed on a blue
continuum.'' The last two properties were chosen to ensure that the
objects in their study had dense regions with  active star
formation. 

It was shown early on that BCD galaxies could not have maintained such
vigorous star formation for long periods of time \citep{searle72}, so
such activity must be episodic. The cause of this activity is
unclear. Unlike spiral galaxies, triggering mechanisms such as spiral
density waves and shear are not present in dwarf galaxies, which tend
to be slow, mostly solid-body rotators. Gravitational instabilities in
the gas, which work well to predict the location of star formation in
large galaxies, seem to work differently in dwarfs, where the average
gas density is typically below the critical \citet{kennicutt89} value by
a factor of 2 or 3, and yet, stars still form.

\citet{he04} showed that
the nature of the \HII\ regions in BCDs are not that different from
those in Ims; they are just crowded more closely together towards the
center in the BCD galaxies. The amount of star formation caused by
pressurized triggering versus spontaneous instabilities is likely
similar as well. So what causes the centralized star formation bursts
that characterize BCDs?  Two possibilities explored in this paper
include an inherent central mass density concentration
\citep{salzer99,vanzee01} or gravitational interactions with nearby
objects, including gas infall/accretion \citep{taylor97, pustilnik01,
  he04, brosch04}. 

Although gravitational interactions are often invoked to explain bursts
of star formation in galaxies, previous studies have shown that not all
the star formation activity in BCDs can be explained by external
gravitational triggers. A plot of current normalized star formation rate
vs.\ distance to the nearest projected (catalogued) galaxy \citep[see Figure
14 in][] {he04} shows no obvious correlation between
presence/distance to companion and star formation activity for Ims,
BCDs, or Sms in a sample of 140 galaxies. This is consistent with other
searches for bright companions \citep{campos91,telles95,telles00}, and
with work done by \citet{brosch04}. This does not rule out possible
triggering by much fainter companions, though, and there is some evidence
for an excess of small/faint companions to BCDs
\citep{taylor97,pustilnik01}, and there are several examples of BCDs
with nearby small galaxies or gas clouds
\citep[e.g.][]{putman98,pustilnik03}.  Such faint companions are
difficult to detect and may have been missed in earlier studies of
BCDs. As pointed out by \citet{he04}, dwarf galaxies are common so ``the
most likely object to perturb any galaxy is a dwarf,'' but the effect of
a small/faint companion on a nearby dwarf galaxy will be more dramatic
than on a much larger system. They suggest that a possible mechanism
for driving mass inflow in dwarfs would be perturbation by a small
companion, resulting in angular momentum
redistribution, and that such interactions could have happened long
ago. Once a low angular momentum system is created, it can continue to
host episodic star formation for long time periods
\citep[e.g.][]{vanzee01}.

There is some evidence for central mass density concentration in BCDs
in both the neutral gas \citep{taylor94,taylor95,vanzee98,simpson00}
and in the optical underlying host galaxy
\citep{papa96a,papa96b,salzer99}.  \Citet{vanzee01} find that the 6
BCDS in their sample have steep rotation curves relative to low
surface brightness dwarfs (LSBDs) with similar luminosities, with
their mass distributions (stars, gas, and dark matter) more centrally
concentrated, and that they are lower angular momentum systems than similar
LSBDs. They suggest that it is just such low angular momentum systems
that are able to collapse into centrally concentrated objects with
high central densities. For a solid body rotation curve, the star
formation threshold density is proportional to the slope of the
rotation curve in a Toomre disk instability analysis
\citep{toomre64}. Hence, a steep rotation curve raises the critical
threshold density for star formation. This delays the onset of star
formation until the gas densities are quite high; and once star
formation does begin, the system is more susceptible to bursting
behavior than a more diffuse (higher angular momentum) system.

Another possible cause for the star formation activity that
characterizes BCDs is the accretion/merger of a small gas cloud. Such
events have been invoked to explain mismatches between the optical and
(gas) kinematic axes in galaxies, twists in the velocity field,
extended/unusual gas morphology \citep[e.g.][]{hunter02},
counter-rotation of the stellar and gas components, and the presence of
stellar ``shell'' features in gas-poor systems
\citep[e.g.][]{coleman04}. Diagnosing the occurrence of a previous
accretion/merger event in dIrr galaxies, however, is complicated by
their low gravitational potentials, large gas fractions, and slow
rotation.  With slower rotation and low gravitational potentials, their
velocity fields are more easily disturbed by the energy injection from
star formation regions, for example.

As part of a project to determine what regulates star formation in
small galaxies \citep{he04,he06}, we have obtained multiwavelength
data ($UBV$ and \ha\ images) for 136 galaxies, with additional data
(e.g. $JHK$, NUV, FUV) for a subset.  These data have recently been
augmented by deep, high resolution interferometric atomic hydrogen
spectral line (\HI) observations of 40 of these galaxies as part of
the LITTLE THINGS (LT) project: Local Irregulars That Trace Luminosity
Extremes: The \HI\ Nearby Galaxy Survey
(http://www.lowell.edu/users/dah/littlethings/).  As part of our
investigation into star formation in dwarf galaxies, we are interested
in looking at ``what goes wrong'' in BCDs---why they have such strong
central star formation occurring. \vii\ is one of a few BCD galaxies in
our sample, and we use it
here as a case study for actively star-forming dwarfs. Because it is
apparently relatively isolated, with no spiral density waves or shear,
it is a good testbed for examining how star formation might be
triggered in active dwarf galaxies. We present the results of an
initial study here; future papers associated with the LT project will
present more in-depth analysis and modeling, particularly of the
kinematics.

Although \vii\ has generally been assumed to be a member of the M81
group, \citet{kara02} state that VII Zw 403, with a distance of 4.4 Mpc,
is behind the M81 group and is being accelerated towards both it and the
Local Group. They find that VII Zw 403 is 1.14 Mpc from M81 and that the
M81 group has a zero-velocity radius of $1.05 \pm 0.07$ Mpc.
\citet{lynds98} and \citet{regina99} determined a distance of 4.5 Mpc
using the tip of the red giant branch from color-magnitude diagrams
based on {\it Hubble Space Telescope} (HST) data. This places it on the
far side of the M81 group. \citet{he04} used the NASA/IPAC Extragalactic
Database (NED)\footnote{http://nedwww.ipac.caltech.edu/} to
search for nearby galaxies in the plane of the sky within 1 Mpc and $\pm
150$ \kms. The only object found (an Im galaxy, KDG 073) had a projected
distance of 900 kpc with a velocity difference of $-32$ \kms.

\citet{loose85} classified \vii\ as an iE galaxy: one with irregular
bright star formation regions near, but not at the center of, a
low-surface brightness envelope. Subsequent observations with HST
confirmed the presence of a smooth elliptical distribution of old,
metal-poor red giant stars with an exponential fall-off in surface
density \citep[and references therein]{lynds98,regina99}. The
scale-length for this evolved population is about 1 kpc \citep{lynds98},
which is normal for BCDs \citep{papa96a}. There are several bright \HII\
regions with young blue stars loosely clustered nearby, and diffuse \ha\
emission as well. Fabry-Perot \ha\ maps \citep{thuan87}
detected little rotation, finding only a chaotic velocity field and a
velocity dispersion of
approximately 30 \kms\ in the diffuse emission, consistent with stellar
winds.

Initial \HI\ inteferometric observations by \citet{thuan04} revealed
that although the \HI\ is centrally concentrated in general in the
galaxy, there are two areas of higher flux density. The velocity
dispersion map showed high values for the region separating the two
peaks, and they interpret the high dispersion in the region between
the peaks as due to a 
lack of flux. As most \HI\ observations of BCDs show smoothly
distributed gas with only one mostly centralized density peak, the
presence of two peaks was intriguing, especially if truly separated by
a region of lower density. To examine the gas morphology and
kinematics in greater detail, and to search for possible companions
and/or evidence of gravitational interactions, we subsequently
obtained high spatial and velocity resolution \HI\ interferometric
data of \vii\ to combine with multiwavelength optical, UV, and IR
data. These data are presented below.

\section{The Observations}

We have collected images of VII Zw 403 at a variety of wavelengths:
H$\alpha$, {\it GALEX} FUV and NUV, ground-based $UBVJ$, and {\it
  Spitzer} 3.6 $\mu$m.  The images are shown in Figure
\ref{fig:combineimages} and observations and data reduction are described
below.

\subsection{Optical Images}

We obtained $UBV$ and \ha\ images of VII Zw 403 with the Perkins 1.8 m
telescope at Lowell Observatory. The $UBV$ images were obtained on
photometric nights with a SITe 2048$\times$2048 CCD binned $2\times2$
for a field of view of 10.4\arcmin. The $V$ and B-band images were
obtained on UT 13 March 1999 and the $U$-band image on 10 April 1999.
The exposure times were $3\times900$ s for $V$, $4\times1800$ s for
$B$, and $3\times1800$ s for $U$. The binned pixel scale was 0\farcs
608, and the seeing on individual and in the final averaged images was
2\farcs1 in $V$, 2\farcs8 in $B$, and 3\farcs 5 in $U$.

The data were reduced using common routines in Image Reduction and
Analysis Facility (IRAF).  Using CCDPROC, we subtracted the electronic
pedestal (sometimes referred to as the ``bias'') using the overscan
strip, and flat-fielded the images using twilight sky flats.  Twilight
flats were found, empirically, to work better than dome flats in
removing large-scale structure in the CCD response.  We used
\citet{landolt92} standard stars in 18 fields over the 3-night
observing run in March 1999 to calibrate the $BV$ photometry and in 13
fields over the 2-night observing run in April 1999 to calibrate the
$U$-band photometry.  The rms of the color terms were 0.015 mag for
$B$, 0.009 mag for $V$, and 0.011 mag for $U$.  The rms of the zero
point terms were 0.015 mag for $B$, 0.088 mag for $V$, and 0.002 mag
for $U$.  The rms of the extinction coefficients were 0.045 mag for
$B$, 0.053 mag for $V$, and 0.005 mag for $U$.  The images in each
filter were averaged (IMCOMBINE) in such a way as to eliminate cosmic
rays while still preserving the photometric integrity of the image.
Specifically, images were compared pixel by pixel for values that
deviated from the average by more than 3 sigma, based on the CCD noise
characteristics. Deviant pixels were then eliminated from the average.

To prepare the images for surface photometry, we edited (IMEDIT)
foreground stars and background galaxies from the final $UBV$ images.
The editing involved replacing a circular or rectangular area with an
interpolation across the region based on the surroundings.  The
$V$-band image was done by hand, and then the cursor log file produced
from that was used to remove objects from the other two filters. In
that way, stars were removed in the same way for each filter.  We then
made a two-dimensional fit to the background (IMSURFIT) using the
non-galaxy part of the image as seen in a  deep display of the
image. We subtracted (IMARITH) this sky image from the object image to
produce a sky-subtracted final image for analysis.
 
We obtained \ha\ images of VII Zw 403 on 4 February 1995 using a TI
$800\times800$ CCD provided to Lowell Observatory by the National
Science Foundation. We used the Ohio State University Fabry-Perot as a
simple 3:1 focal reducer since we were not interested in narrow-band
spectral information and the \ha\ filter had a higher efficiency than
the Fabry-Perot etalon.  The \ha\ filter has a central wavelength of
6566 \AA\ and a FWHM of 32 \AA.  The contribution from stellar
continuum was determined using an off-band filter centered at 6440
\AA\ with a FWHM of 95 \AA.  The \ha\ observations consisted of two
1300 s images, preceded and followed by a 600 s off-band observation.
Given the FWHM of the filters, this approximately matched the S/N of
the on and off-band filters.  The pixel scale was 0\farcs49, and the
seeing on the individual and final image is 1\farcs7.  The electronic
pedestal (``bias'') was subtracted using the overscan strip, and the
images were flat-fielded using dome flats.


We shifted, scaled, and subtracted the off-band image from the \ha\
image to remove the stellar continuum and leave only \ha\ nebular
emission.  We calibrated the \ha\ flux using observations of 5
spectrophotometric standard stars \citep{oke74,stone77}  and
observations of the NGC 2363 nebula \citep{kennicutt80} taken over the
course of the 6-night observing run.  These two types of calibrations
agreed with each other to within 4\%.  The \ha\ flux has been corrected for
the shift of the bandpass with temperature. The central wavelength
shifts blueward by 0.018 nm for each degree of temperature decrease
from 20$\arcdeg$ C, and so this needs to be considered in calibrating
the flux through these narrow-band filters as it sometimes moves the
object emission line off the peak of the transmission function.  The
\ha\ flux has also been corrected for contamination by the
[NII]$\lambda$6584 emission line in the bandpass (a 1\% correction).

\subsection{Near-IR Image}

We obtained the $J$-band image of VII Zw 403 with the Perkins 1.8 m
telescope at Lowell Observatory with the Ohio State Infrared
Imager-Spectrograph \citep[OSIRIS][]{depoy93} on the photometric
night of UT 2 June 1996.  The instrument consisted of four mosaiced
detectors to create a total array of 256$\times$256 pixels with a
field of view of 6.4\arcmin.  The telescope was nodded to place VII Zw
403 first in one half of the array and then in the other half, so that
sky could also be measured in the half without the galaxy. The galaxy
was offset a few arcseconds between images in order to allow removal
of hot pixels when the frames were averaged later.  The image of VII
Zw 403 is a combination of 48 exposures of 26 s each.  The pixel scale
was 1\farcs50, and the seeing of the final image was 3\farcs2.

The data were reduced in IRAF. Zero second bias frames were subtracted
from all of the data, the image pixel values were corrected for
non-linearity effects, and pixel to pixel sensitivity variations were
removed using observations of a white screen in the dome.  The
non-linearity corrections were determined using a series of dome flats
of increasing exposure time, to 29 seconds in steps of 1 second,
sandwiched between darks.

Eight UKIRT standard stars from the list compiled by S.\ Courteau
(1996, private communication) were used to calibrate the photometry.
Each star was observed 4 times with the star placed in each quadrant
in turn, and sky for a specific observation was taken from the
observations with the star in the other quadrants.  The rms of the
zero point of the $J$-band calibration was 0.12 mag and of the
extinction term was 0.08 mag.  The California Institute of Technology
(CIT) photometric system is 8\% bluer in $J-H$ and 4\% bluer in $H-K$
compared to the UKIRT system.  The star forming region of VII Zw 403
shows up faintly on the Two Micron All Sky Survey
\citep[2MASS][]{skrutskie06} $J$-band image:
aJ\_asky\_990319n0730056.fits.  On the 
2MASS image, the flux in a 22\arcsec\ radius aperture centered on the
star-forming regions has a magnitude of 13.73$\pm$0.06. Our $J$-band
image gives 13.71$\pm$0.01. In spite of the difference in calibration
systems (2MASS defines their own system), the magnitudes are the same
within the uncertainties.

\subsection{Ultraviolet Data}

We obtained near-UV (NUV) and far-UV (FUV) images of VII Zw 403 from
the {\it Galaxy Evolution Explorer} ({\it GALEX})
archives\footnote{Based on observations made with the NASA {\it Galaxy
Evolution Explorer} \citep{Martin_etal05}.  {\it GALEX} is
operated for NASA by the California Institute of Technology under
NASA contract NAS5-98034.}.  The images were part of the Nearby
Galaxy Survey.  The FUV filter has a bandpass of 1350--1750 \AA, an
effective wavelength of 1516 \AA, and a resolution of 4\farcs0. The
FUV exposure time was 1617 s. The NUV filter has a bandpass of
1750--2800 \AA, an effective wavelength of 2267 \AA, and a resolution
of 5\farcs 6. The NUV integration time was 1479 s.  The images were
processed through the {\it GALEX} GR2/3 pipeline and were retrieved as
final intensity maps with a 1\farcs 5 pixel scale.  The {\it GALEX}
field of view is a circle with 1\fdg2 diameter, and we extracted a
portion around VII Zw 403.  We subtracted (IMEDIT) or masked
foreground stars and background galaxies, and subtracted a low order
two-dimensional fit to the background (IMSURFIT).  The decision of
what objects did not belong to the galaxy and should be removed was
based on morphology (extended or stellar profiles), color, and
similarity with other objects in the field well beyond the galaxy
itself.

\subsection{3.6 $\mu$m Data}

For completeness and to facilitate comparison, we include the 3.6
$\mu$m image taken of VII Zw 403 with the InfraRed Array Camera
\citep[IRAC][]{fazio04} on the {\it Spitzer Space Telescope} ({\it
  Spitzer}).  The data were obtained from the {\it Spitzer} archive
and reduced with the Basic Calibrated Data pipeline.  The full
treatment of the data is described by \citet{hunter06}.

\subsection{\HI\ Data}

21 cm spectral line observations of the galaxy were made using the B, C,
and D configurations of the NRAO\footnote{The National Radio Astronomy
  Observatory is a facility of the National Science Foundation operated
  under cooperative agreement by Associated Universities, Inc.}. Very Large Array (VLA).
The B and D-configuration data were taken by Deidre Hunter; the C
configuration by Trinh X.\ Thuan (proposal code AH0453), who kindly
supplied us with these observations. The D configuration, with its short
spacings, provides high sensitivity but a relatively low spatial
resolution of $\sim$ 1\arcmin. The C and B configurations, with longer
baselines, provide higher resolution (nominally 12\farcs5 and 4\arcsec,
respectively) but lower surface brightness sensitivity.  After editing
and calibration (discussed below), the three data sets were combined to
produce images with both high sensitivity and high angular resolution.

Time on source for the C and D configuration observations was
approximately 4 hours each; for the B, approximately 9 hours.  Observations
for all configurations were made using a correlator setting using 2IF
mode with Hanning smoothing with a total bandwidth of 1.56 MHz. 
This results in 128 channels with a channel separation of 12.2 kHz (2.6
\kms). For spectral line data that are taken using on-line Hanning
smoothing, the velocity resolution is equal to the channel separation,
and so is also 2.6 \kms. Observational information is provided in
Table~\ref{tab:HIObs}.

The D configuration observations were set up to be consistent with the
previously existing C configuration observations, so the two sets of
observations could be combined after calibration for analysis. The
central observing frequencies and velocities were the same, other than
the usual changes in sky frequency due to the day, time, and location of
the sources on the sky. However, the B configuration observations were
taken during the beginning of the transition of the VLA to the Extended
VLA (EVLA); so doppler tracking while observing was not available. The
observations were done at a fixed frequency which corresponded with the
central velocity ($-60$ \kms) from the C and D observations to within 3
or 4 channels (approximately 12 kHz). Before the B data were combined
with the other two data sets (see below), the {\sc aips} task {\sc cvel}
was used to shift the spectra in the B data set to correct for the
Earth's rotation and motion. Careful comparison of the resulting
spectrum after applying {\sc cvel} with those from the C and D data
ensured that all three data sets were now consistent in frequency and
velocity space.

\subsubsection{Calibration}\label{sec:calibration}

Calibration of all three data sets was performed independently, using
the standard routines in {\sc aips}. For the D configuration data, the
flux and bandpass calibrator was 1411+522 (3C 295; all sources are in
J2000) and the phase calibrator was 1435+760 (J2000).  The C
configuration observations used 1044+809 for phase calibration and 1331+305
(3C 286) for the flux and bandpass calibration. The phase calibrator for
the B configuration data was the same as for the D (1435+760), and the
flux and bandpass calibrator was 0137+331 (3C 48).

The data reduction process was
complicated by the presence of both absorption and emission 
from the Milky Way as \vii's systemic velocity is $\sim -100$ \kms. The
single-dish spectrum (Figure~\ref{fig:tmspec}) from the Greenbank 91m telescope \citep{tm81}
 shows a 
broad absorption feature from $\sim -55$ to $-30$ \kms, with a strong
emission feature from $-$30 to $\sim +5$ \kms. That emission feature shows
a dip due to a narrow absorption line at a velocity of about 0
\kms. 

To reduce the amount of data potentially contaminated by Galactic \HI\ in the
D configuration calibrator data, a pseudo-continuum ``channel 0''
dataset was constructed by averaging together the inner 50\% of the line
data (normally, the inner 75\% is used). This probably did not remove
all the emission from the Milky Way, but was the best compromise given
the situation. Because these were daytime
observations, solar contamination was also present on the shortest
baselines for the calibrators as well as for the source data. The
affected short baselines for the calibrator data were not used during
the calibration by the application of an appropriately chosen
``uvrange'' (only baselines longer than 0.55 k$lambda$ were used).  Once
a satisfactory calibration was achieved, it was 
applied to the source line data. The bandpass calibrator was only
affected by Milky Way \HI\ in a few channels, and as those hadn't stood
out as unusual in the previous solutions, the bandpass calibration was
done, again applying the uvrange limit (0.55 k$\lambda$) for the solar
contamination. The continuum emission was removed using {\sc uvlin},
which does a linear fit to specified (line-free) channels and then
subtracts the fitted baseline from the spectral line channels. The
channels used for the fit were chosen to be free of both line emission
and Galactic contamination. 

For the C configuration observations, the continuum source 1331+305
(J2000) was observed for use as a flux and bandpass calibrator; the
nearby continuum source 1044+809 (J2000) was observed for phase
calibration. There was little Galactic or solar contamination present in
these data. For the C data, calibration proceeded normally. During
editing of the channel 0 source data, amplitudes above 0.4 Jy were
clipped to remove anomalously high amplitude visibilities on the
shortest baselines during the beginning of the observation run. The
editing and calibration tables were copied over to the line data, and a
dirty map of the source data was made to check for interference and look
for line-free channels to use for continuum subtraction. Continuum
subtraction using the line and inteference-free channels thus identified
was done in the {\it u-v} plane using the task {\sc uvlsf}, which works
similarly to {\sc uvlin}.

The B configuration observations used 1331+305 (J2000) for flux and
bandpass calibration, and 1435+760 (J2000) for phase calibration. The
data from four antennas were flagged as unusable during the data
collection process due to EVLA testing; four more
antennas were not in use during the observations. During calibration,
several large phase jumps in the phase calibrator data, on the order of
100 degrees, were observed for some of the antennas. All of these jumps
occurred between scans near the end of the observing run. Changing the
reference antenna for the calibration did not change the number or
magnitude of the problematic jumps. After further consideration and
testing, the galaxy source data from the scans that were in between the
calibrator scans with phase jumps larger than 40 degrees were removed by
flagging. Careful examination of the resulting images revealed no
artifacts that would indicate phase errors with the
data, such as asymmetric stripes, rings, or other odd-looking
features. Self-calibration was also tried, but the resulting images were 
essentially identical to those from the edited data. Results here are
from the edited data that were not self-calibrated. As with the D data,
continuum subtraction was done in the {\it u-v} plane using {\sc
uvlin}. There was no apparent contamination from Milky Way emission
visible in the B configuration data.

To check the calibration and data quality, the data from all three
configurations were imaged separately using the task {\sc imagr}, which
uses a fast-fourier transform to create a cube of images, one image for
each frequency (velocity) channel. During imaging, the effects of
side-lobes of the
non-gaussian features of the ``dirty'' beam were removed using the {\sc
clean} algorithm (with a flux cut-off equal to the r.m.s.\ in a
line-free channel). With the exception of the Galactic contamination
obvious in the D configuration data (and  faintly in a few channels
of the C data), all three data sets were deemed acceptable.

A spectrum from our D configuration data (uniform weighting, with a {\sc
robust} factor of $+1$)\footnote{The {\sc robust} parameter allows the
user to choose a weighting scheme between fully uniform ($-5$) and
fully natural ($+5$). This produces images with either higher spatial
resolution but poorer signal-to-noise (fully uniform) or lower spatial
resolution but higher signal-to-noise (fully natural).} is shown in
Figure~\ref{fig:dispec}.  Some of the Galactic contamination is strongly
visible around 0 \kms, which doesn't overlap with \vii's emission,
visible around $-$100 \kms.  Many of the same features are present here
as are in the \citet{tm81} single-dish spectrum, although at different
levels. Presumably this is due to the spatial scale of the features; the
D configuration is ``blind'' to structures larger than about 15\arcmin\
per spectral channel. 

The D configuration data were then reimaged using natural weighting to
emphasize faint emission; the spatial resolution is $72\farcs9
\times53\farcs6$ and the $1\sigma$ noise in a single channel is 0.93 mJy 
beam$^{-1}$ (see Table~\ref{tab:HIObs}).  Our integrated flux from the D
configuration data is 11.5 Jy \kms, which is 25\% lower than previous
single-dish measurements. (We did not correct for the sensitivity fall-off
of the primary beam, as the maximum extent of \HI\ emission was
approximately 6\arcmin, which is well within the 30\arcmin\ full-width at
half-maximum (FWHM) of the primary beam at 21cm.) Estimates from single
dish observations are 
14.6 \citep{hr86}, 15.8 \citep{tm81}, and 16.2 Jy \kms\ \citep{hr89}. We
checked our flux calibration by comparing the fluxes of continuum
sources from our data to the catalogued fluxes listed in the NRAO VLA
Sky Survey (NVSS\footnote{http://www.cv.nrao.edu/nvss/};
\citet{condon98}). There are two catalogued continuum sources close
(3.7\arcmin\ and 3.5\arcmin) to the galaxy; our measured fluxes are 79\%
and 76\% of the NVSS fluxes, respectively. If we adjust our D
configuration integrated flux by 20\%, this still only comes to 13.8 Jy
\kms. A comparison of a spectrum through the D configuration data cube
restricted in area to the spatial extent of the galaxy on the sky is
shown in Figure \ref{fig:single}, along with the single-dish spectrum
from \citet{tm81}. Presumably, all the single-dish integrated fluxes
include the second emission peak located at about $-70$ \kms. We have
determined that this secondary emission peak is almost certainly from
the Milky Way, not \vii\ itself. Comparing just the emission from \vii,
it is clear that our observations have detected all of the flux seen in
the single-dish spectrum.

This is a different conclusion than that reached by
\citet{thuan04}. They present the VLA C configuration data that we show
here, but did not have the D configuration data and so were not aware of
the Galactic emission in the field of view. They also present a
comparison of their C data spectrum to the original \citet{tm81}
single-dish spectrum. Their C spectrum resembles our D spectrum without
the Milky Way emission; indeed their total integrated flux from the C
configuration data is 11.0 Jy \kms, which is close to our 11.5 Jy \kms\
value. However, without knowing that the single-dish flux included the
Galactic emission as well, they were forced to conclude that their VLA
data were missing some (presumably large-scale) emission from \vii.

\subsubsection{Imaging}

Channel maps from part of the D configuration data cube are shown in
Figure~\ref{fig:dchanmaps}. The Galactic contamination is clearly
visible in the channels from $-23.9$ to $-78.0$ \kms. Several of these
channels also contain emission from \vii. Before integrating the cube to
produce the moment maps showing the integrated flux, velocity field, and
velocity dispersion, the affected channels were selectively blanked to
remove the Galactic emission, leaving only the small region in the
center of each field that contained emission from \vii. This partially
blanked cube was then integrated in velocity using the channels from
$-60.0$ to $-147.5$ \kms\ with the {\sc aips} task {\sc momnt}. To avoid
integrating up noise as well as signal, {\sc momnt} creates a mask cube
smoothed in space (by convolving with a gaussian twice the size of the
beam) and velocity (by hanning smoothing); pixels falling below the
specified flux 
cutoff in the mask cube are then blanked in the original cube prior to
integration. For this dataset, a flux cutoff of twice the
single-channel r.m.s.\ was used. These moment maps were used for the
analysis discussed in \S~\ref{sec:radio}.

To produce a data set with good sensitivity and enhanced but still
moderate resolution, the data from the C and D configurations were
combined in the {\it u-v} plane using the {\sc aips} task {\sc dbcon}
and then imaged using the {\sc aips} task {\sc imagr} with a {\sc
robust} weighting factor of $+1$ (we will refer to this cube as the
\lres\ cube). This weighting was chosen based on tests to determine the
best balance of resolution and flux recovery (sensitivity) for the
purposes of velocity field analysis, primarily.  Although the Galactic
contamination wasn't as strongly present as in the D-configuration-only
data, it was still present at high enough levels to warrant concern. We
therefore blanked everything but the emission from \vii\ (in the center
of the field) in the eight affected channels before integrating. A flux
cutoff of twice the single channel r.m.s (see Table~\ref{tab:BCDdata})
was used to produce the combined data moment maps. This data set was
used for the velocity analysis and \HI\ surface density analyses, below.

The data from all three configurations were also combined in the {\it
u-v} plane and imaged. Galactic contamination was not detectable in any
of the resulting data cubes, so no extra blanking was necessary. We
tested a variety of {\sc robust} weighting factors while imaging to
determine the best values to provide the desired resolutions while
recovering the most flux possible for each resolution. After extensive
testing and comparison, we created three
data cubes with varying sensitivities and resolutions: a {\sc robust} of
0 produced a medium resolution data set (\mres) which was used to examine the
morphology of the integrated flux (the first moment map); a {\sc robust}
of +1 produced a lower resolution data set (\bcdone) used to analyze the
velocity dispersion in the gas (the second moment map); and a {\sc robust} of
$-1$ was used to create a high resolution (\hres) image of the
integrated flux in the highest density regions of the galaxy. Details
(resolutions, noise levels) are given in Table~\ref{tab:BCDdata}. 

\section{The Optical View of the Galaxy}\label{sec:optical}

The images of VII Zw 403 are shown in Figure \ref{fig:combineimages}.  The
galaxy is elliptical-shaped with a moderate degree of ellipticity
implying a relatively high inclination angle.  At outer $V$-band
isophote levels VII Zw 403 is quite symmetrical.  However, at
intermediate isophotes (radius 40--60\arcsec) there is an extension of
stars to the north that is not mirrored to the south.  This type of
feature is seen in other irregulars as well, such as NGC 2366 \citep{hew01}.


\subsection{Broad-band Surface Photometry}

From the $V$-band image, we determined the morphological center of the
galaxy, $b/a$, and position angle of the major axis from a contour
plot of the image block-averaged by 5 to increase the signal-to-noise
in the outer parts. We used an outer contour corresponding to a
$\mu_V$, not corrected for reddening, of 27.2 mag arcsec$^{-2}$.  The
major axis and position angle were determined to be the long line of
symmetry around which the galaxy could be folded. The center was the
mid-point bisecting the isophote along the major and minor axes.

The center of the galaxy is at 11$^{\rm h}$ 24$^{\rm m}$ 33\farcs4, 79\arcdeg\
16\arcmin\ 10\arcsec\ (B1950), with an uncertainty of 2\arcsec\ in
each direction.  We also determined the center of the galaxy from {\it
Spitzer} IRAC 3.6 $\mu$m images \citep{hunter06}, with an estimated
uncertainty of 3\arcsec.  The 3.6 $\mu$m center is offset from that of
the $V$-band by 4\arcsec\ W and 3\arcsec\ N, and this then more
realistically describes the uncertainty in the position of the center
of the stellar body.  Below we will derive the \HI\ kinematic center
as well. That center is at 11$^{\rm h}$ 24$^{\rm m}$ 35\fs1$\pm10\fs3$,
79\arcdeg\ 
16\arcmin\ 18$\pm$27\arcsec\ (B1950).  The \HI\ kinematic center is
offset from that of the $V$-band by 5\arcsec\ E and 8\arcsec\ N, well
within the large error bars of the \HI\ kinematic center.  For the
surface photometry we used the $V$-band morphological center.  We
measured the minor-to-major axis ratio $b/a$ on the $V$-band image to
be 0.49. Dwarf irregular galaxies are considered to be thicker than
spiral disks and an intrinsic ($b/a$)$_0$ of 0.3 is suggested
(\citealt{hodge66}; \citealt{vandenbergh88}; \citealp[but
see][]{sung98}).  This $b/a$ then implies an inclination of 66\arcdeg\
for VII Zw 403.  We measured the position angle of the major axis to
be $-11$\arcdeg.

The surface photometry (ELLIPSE in STSDAS) used 10 ellipses starting
at a semi-major axis of 9\arcsec\ and progressing to 91\arcsec\ in
steps of 9\farcs1.   
All of the other images were geometrically transformed (GEOMAP,
GEOTRAN) to match the $V$-band image and elliptical surface photometry
was done identically on all images. The geometric transformation used
stars identified in both the $V$-band image and the target image to
determine the surface distortions necessary to make the target image
have the same size, pixel scale, and orientation of the $V$-band
image.

The surface photometry is collected in Figure \ref{fig:sb}.  The
surface photometry and colors have been corrected for reddening using
a total E($B-V$)$_t$=E($B-V$)$_f$$+$0.05, where the foreground
reddening E($B-V$)$_f$ is 0.023 \citep{bh84}.  An internal reddening
of 0.05 in E($B-V$) was used to be consistent with Balmer decrement
measurements of \HII\ regions in dIm galaxies.  \citet{hunter99} found
an average E($B-V$) of 0.1 for \HII\ regions in 39 dIm 
galaxies. The stars outside of \HII\ regions were taken to have a
reddening of half of this value.  We used Burstein and Heiles values
for the foreground reddening rather than those of \citet{schlegel98}
because the latter values are systematically higher than the 
former and some are inconsistent with Balmer decrement measurements.
Our value of the reddening of the stars in VII Zw 403 is close to that
of \citet{lynds98} who obtained an E($B-V$) of 0.04--0.08 for stars
and stellar associations from analysis of color-magnitude diagrams.
We use the reddening law of \citet{cardelli89} and
$A_V$/E($B-V$)$=$3.1 for the optical, near-IR, and FUV reddening
corrections.  The NUV reddening is affected by the 2175 \AA\ bump. For
the NUV, we have adopted the reddening formula of \citet{wyder07}.

We determined $R_{25}$, the radius at a $B$-band surface brightness of 25
mag arcsec$^{-2}$, to be 36\farcs5 which is 0.78 kpc at the galaxy.
Our radius is 10\% smaller than that given by \citet{rc3}.  The Holmberg
radius, $R_{\rm{H}}$, originally defined to a photographic surface brightness,
is measured at an equivalent $B$-band surface brightness $\mu_B = 26.7 -
0.149(B-V)$. For our measured ($B-V$)$_0$ of 0.3, the Holmberg radius is
determined at a $\mu_{B_0}$ of 26.66 magnitudes arcsec$^{-2}$. We find
that $R_H$ is 69\arcsec\ which is 1.47 kpc at the galaxy. Both $R_{25}$
and $R_H$ are determined from the reddening corrected surface
photometry. Integrated properties of \vii\ are collected in Table
\ref{tab:global}.

We fit an exponential disk to the broad-band photometry beyond 0.2
kpc, and the fits to the $V$-band, 3.6 $\mu$m, and NUV data are shown
in the upper panel of Figure \ref{fig:sb}.  The outer disk of \vii\ is
fit well with an exponential disk profile having a central surface
brightness $\mu_0^V$ of 22.94$\pm$0.09 magnitudes arcsec$^{-2}$ and a
disk scale length $R_D^V$ of 0.50$\pm$0.01 kpc.  Interior to a radius
of 1.0 kpc, the surface brightness profile is steeper, although it too
is fit well with an exponential disk with a disk scale length of
0.37$\pm$0.01 kpc.  The 3.6 $\mu$m surface brightness profile follows
the slope of the $V$-band reasonably well while the NUV profile is
considerably steeper. The similarity between the $V$ and the 3.6
$\mu$m profiles implies that the star formation as a function of
radius integrated over a 1-Gyr time scale is the same as that
integrated over the lifetime of the galaxy. The steeper NUV profile
suggests that star formation over the last 200 Myrs has been
concentrated to the center of the galaxy.

Surface brightness profiles of BCDs are generally found to contain
multiple components with the outer component representing the galaxy
underlying the intense current star formation
\citep{papa96b,cairos03,noeske03}.  The central portions of these
galaxies, on the other hand, are dominated by the current star
formation, and the profiles are often steeper or more complex there.
\citet{papa96a} found that the scale-length of the outer disks of BCDs
are smaller, by half on average, than those of Im galaxies.  However,
\citet{he06} find for a sample of 94 Im systems and 24 BCDs that the
scale-lengths of the BCDs are only slightly offset (in the sense of
shorter scale-length at comparable $M_V$) from those of most Im
galaxies. It is the higher scale-length tail in the number
distribution that is more drastically reduced for their BCD sample
compared to dIms.  Compared to these samples, \vii's scale-length lies
at the peak in the distribution of dIms and BCDs.

Integrated colors of \vii\ are given in Table \ref{tab:global}.  In a
$UBV$ color-color diagram for a large sample of Im, BCD, and Sm galaxies
\citep{he06}, \vii\ lies at the blue side of the range, clearly
dominated by a young stellar population.  On the other hand, the
integrated ($V-J$)$_0$ places \vii\ in the middle of the range shown
by Im and BCDs in that color \citep{he06}.

Azimuthally-averaged radial profiles of the colors are shown in Figure
\ref{fig:sb}.  In most Im galaxies the azimuthally-averaged colors are
generally constant over the disk of the galaxy; irregular galaxies do
not usually have color gradients. \vii, however, does show a gradient
towards redder colors in the outer part of the galaxy.  $(B-V)_0$ and
$(U-B)_0$ track each other and change by about 0.5 mag from the center
into the outer galaxy, beyond the Holmberg radius. We can only trace
$(V-J)_0$ to 1.1 kpc, but it is approximately constant over this
radial range. $(FUV-NUV)_0$ also is roughly constant throughout the
optical galaxy to $R_{25}$.  This trend in the colors is due to the
fact that star formation is concentrated to the center of the
galaxy. Stellar populations becoming redder as one goes outwards from
the center is seen in many BCDs \citep{papa96b,cairos03,noeske03}.

We have modeled the UV and optical colors of VIIZw403's annular
surface photometry using population synthesis fitting. These models
use  simple evolutionary histories and the stellar
populations library of \citet{bruzual03}. Details can be found in
\citet{hunter10}. We find that the colors of the central region
($<$0.5 kpc) is dominated by a stellar population with an age of about
400$\pm$200 Myrs, while the colors beyond that radius are fit with
stellar populations of 1-3 Gyr of age.

\subsection{Star-Forming Regions}

\begin{figure}
\epsscale{0.5}
\plotone{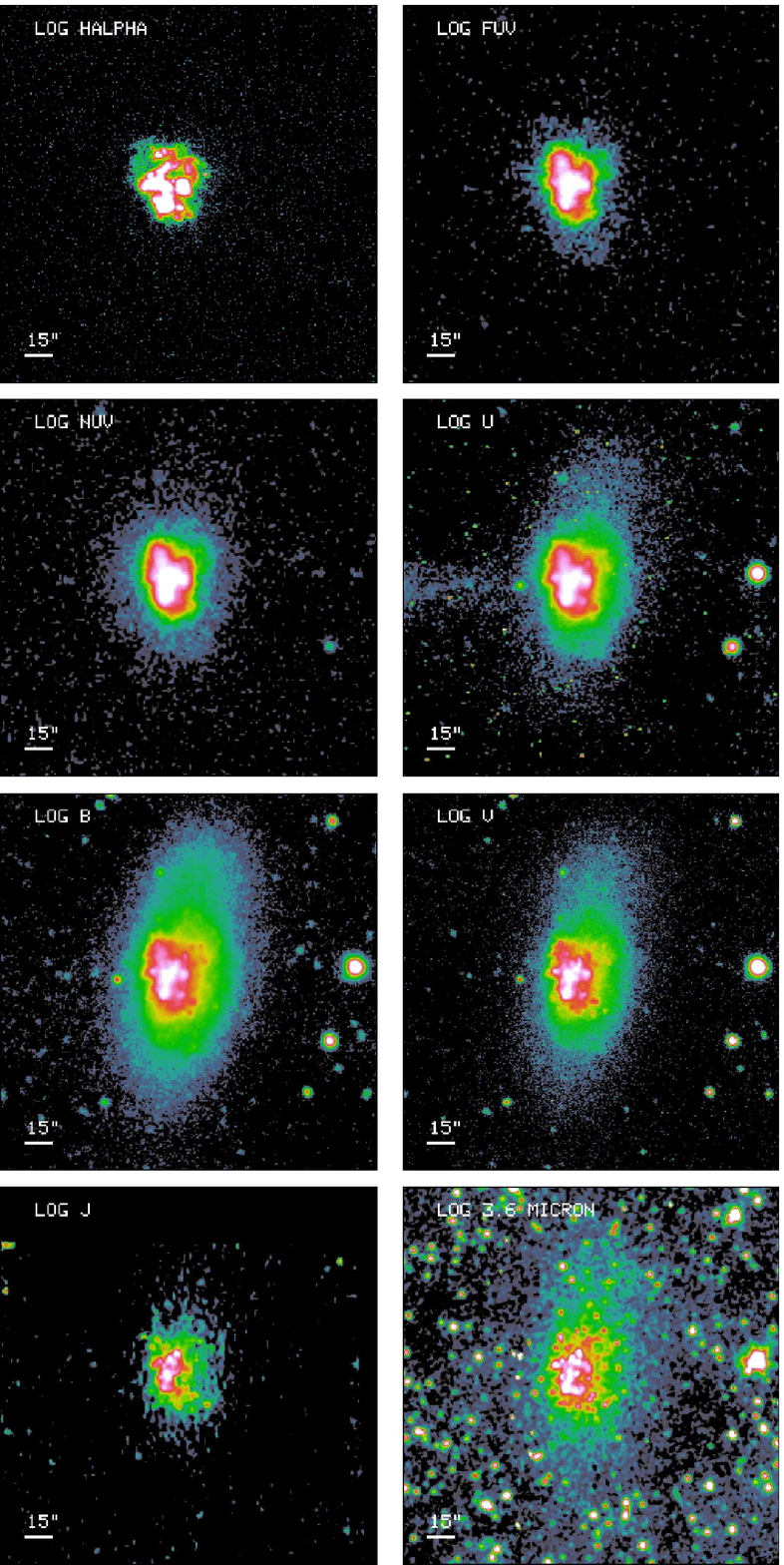}
\caption{Images of VII Zw 403 in various passbands. The images are
  shown as the logarithm in order to allow the inner and outer
  structure to be viewed simultaneously. The images are also
  geometrically transformed to match the scale and orientation of the
  $V$ image, and the same field of view is shown in each plot. North
  is up and East to the left.
\label{fig:combineimages}
}
\end{figure}

\begin{figure}
\epsscale{1.0}
\plotone{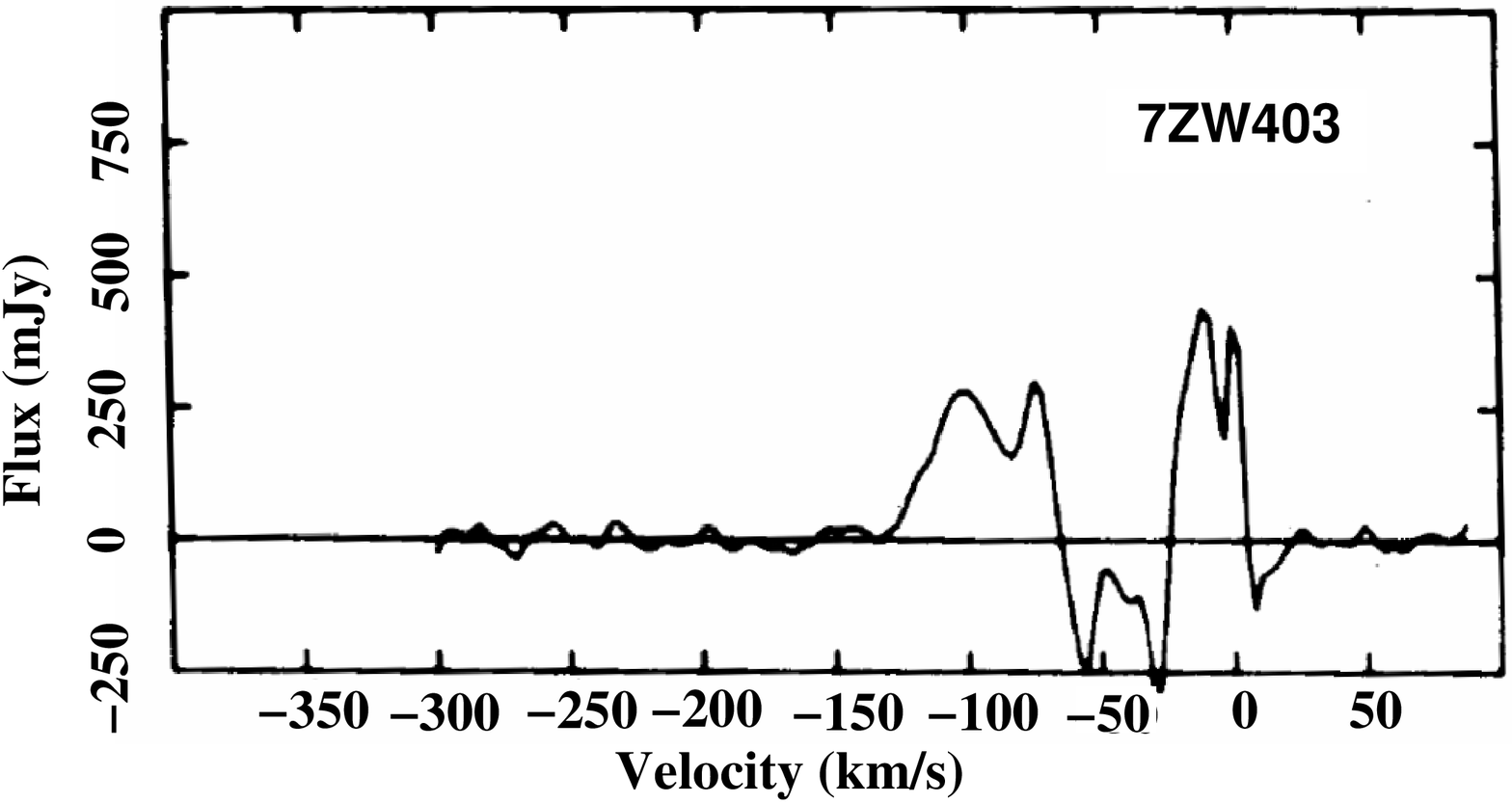}
\caption{Single-dish spectrum from \protect\citet{tm81}.
\label{fig:tmspec}
}
\end{figure}

\begin{figure}
\epsscale{1.0}
\plotone{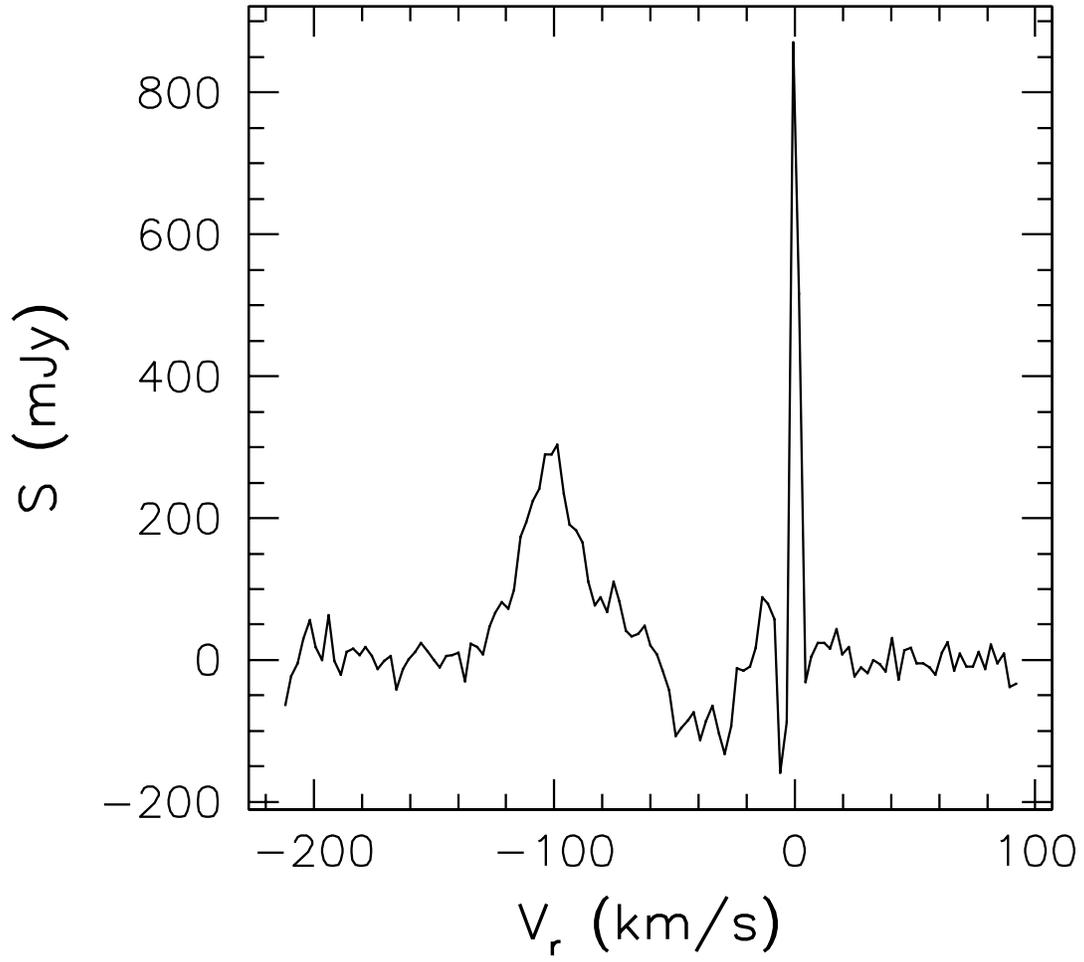}
\caption{Summed flux in each channel from the uniformly-weighted D
  configuration data. The strong spike and absorption features near 0
  \protect\kms, as well as some of the emission from approximately $-55$
  to $-80$ \protect\kms\ are due to Galactic \protect\HI.
\label{fig:dispec}
}
\end{figure}

\begin{figure}
\epsscale{1.0}
\plotone{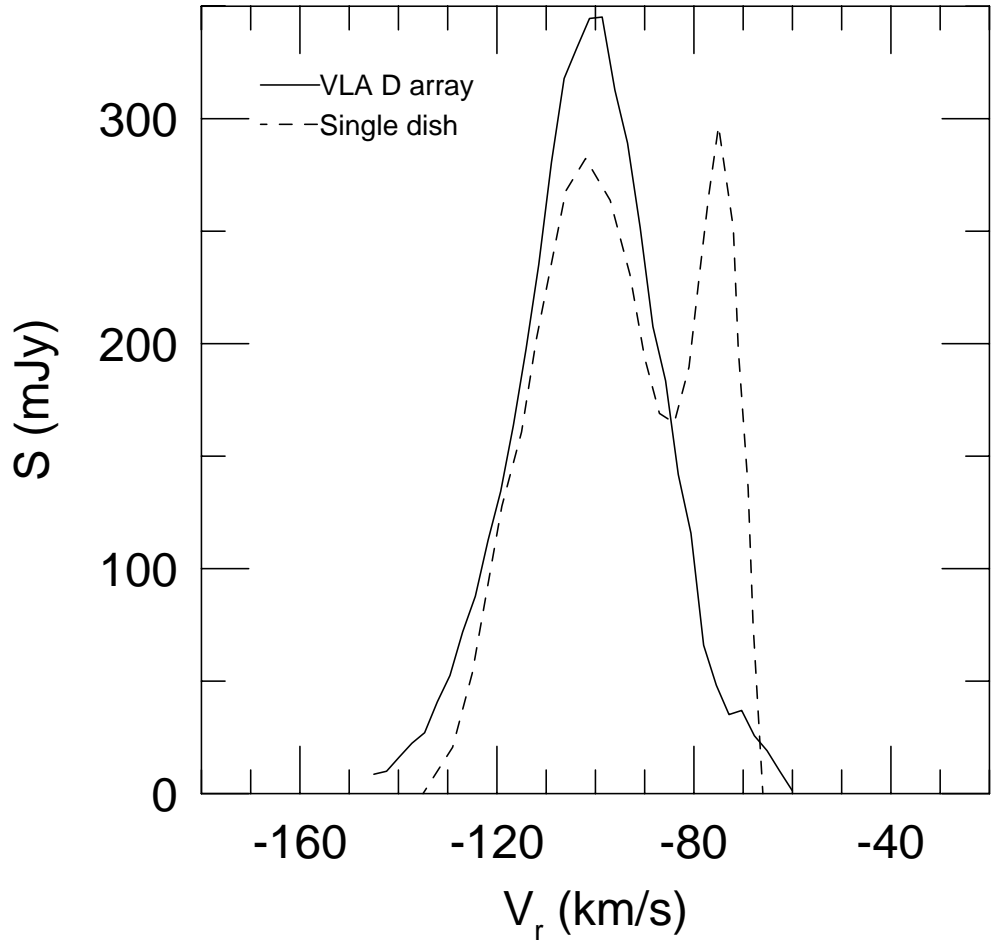}
\caption{Summed flux in each channel from the naturally-weighted D
  configuration data (solid line) is plotted along with the single-dish
  detection by \protect\citet{tm81} (dotted line). The emission located around
  $-70$ \protect\kms\ is foreground \protect\HI\ from the Milky Way.
\label{fig:single}
}
\end{figure}

\begin{figure}
\epsscale{1.0}
\plotone{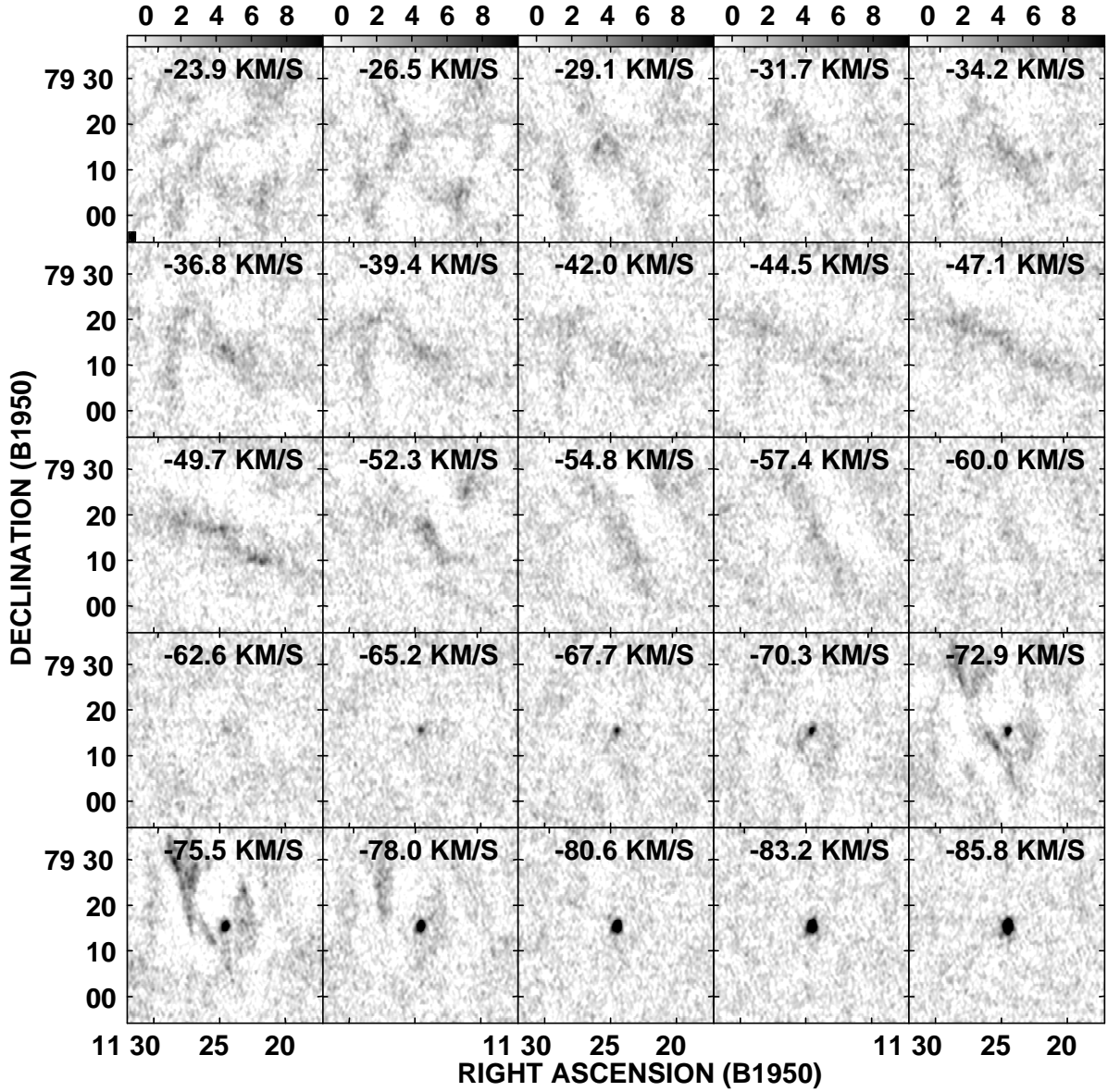}
\caption{Channel maps from part of the D configuration data
  cube. Contaminating emission from the Milky Way is visible in the
  velocity range from $-23.9$ to $-80.6$ \protect\kms. The strong
  concentrated emission in the center of the field is from
  \protect\vii.
\label{fig:dchanmaps}
}
\end{figure}

\begin{figure}
\epsscale{0.75}
\plotone{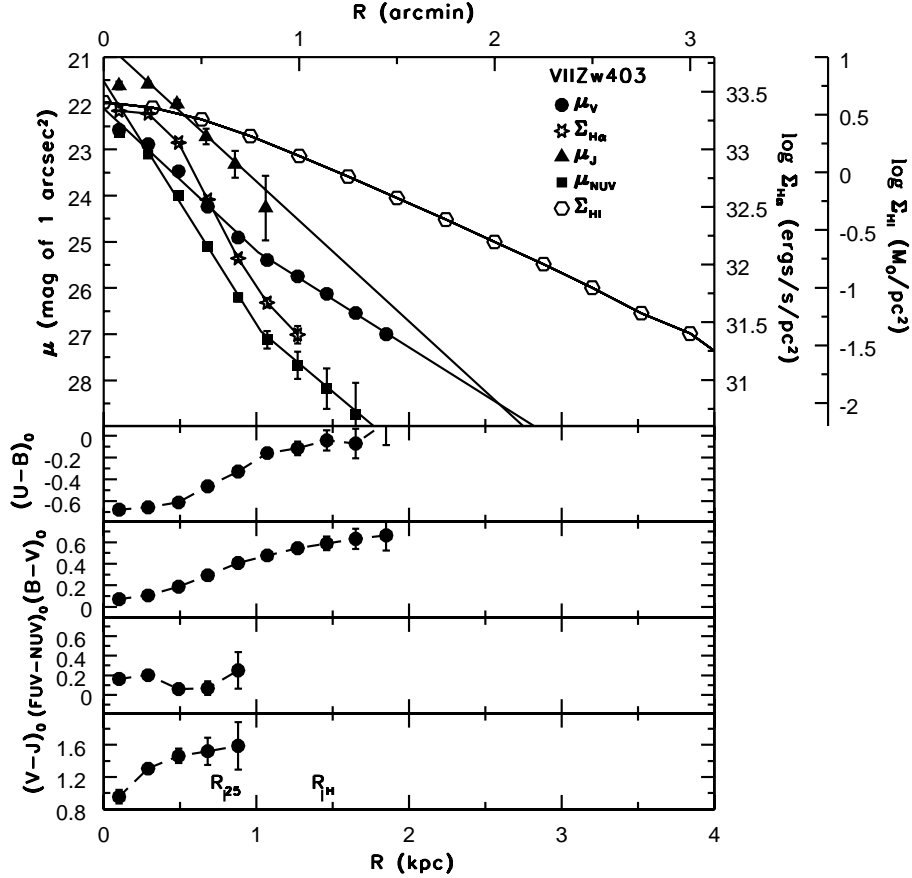}
\caption{Azimuthally-averaged $V$, H$\alpha$, NUV, 3.6 $\mu$m,
$(U-B)_0$, $(B-V)_0$, $(FUV-NUV)_0$, and $(V-J)_0$ surface photometry
and \HI\ surface density of \vii, using the $V$-band morphological
center of the galaxy.  The $UBVJ$ and NUV photometry are corrected for
reddening as described in the text. The exponential fits to the $V$, 3.6
$\mu$m, and 
NUV-band surface brightness profiles are shown as solid lines in the top
panel.  These profiles are fit with two exponentials; the outer one
shallower than the inner.   The H$\alpha$ luminosity is
proportional to the star formation rate.  The scales for
$\Sigma_{H\alpha}$ and for $\Sigma_{HI}$ have been set to cover the same
logarithmic interval as the broad-band surface photometry covers in
magnitudes and are shown on the right y-axis.
\label{fig:sb}
}
\end{figure}

H$\alpha$ traces star formation over the past 10 Myr, and Figure
\ref{fig:combineimages} shows that most of the current star formation in VII Zw
403 is found in one  large complex located 6\arcsec\ east and
2\arcsec\ north of the center of the galaxy, only 130 pc from the
nominal center determined from the outer $V$-band isophote.  This type
of central concentration of star formation is often seen in Blue
Compact Dwarfs, and is probably responsible for VII Zw 403's
classification as a BCD.

We show the azimuthally-averaged \ha\ surface brightness in Figure
\ref{fig:sb}.  Although the current star formation in VII Zw 403 is
contained in a giant \HII\ complex at the galaxy center (the H$\alpha$
emission seen in Figure \ref{fig:combineimages}), because the \HII\ complex is
large, \ha\ emission appears in elliptical annuli out to a radius of
1.3 kpc.  The \ha\ surface brightness drops off faster than the
starlight in the $V$-band and 3.6 $\mu$m image, dropping only a little
faster beyond 0.75 kpc and ending at 1.3 kpc.

The FUV and NUV passbands measure the light directly from stars and
are dominated by young massive stars formed over the past 200 Myr.
Thus, the UV is sensitive to star formation over a longer timescale
than \ha.  In Figure \ref{fig:sb} the UV profile, however, follows the
\ha\ profile closely. Thus, the ultraviolet surface brightness profile
is dominated by the same central star-forming event as we see in \ha\
(see Figure \ref{fig:combineimages}).

\begin{figure}
\epsscale{1}
\plotone{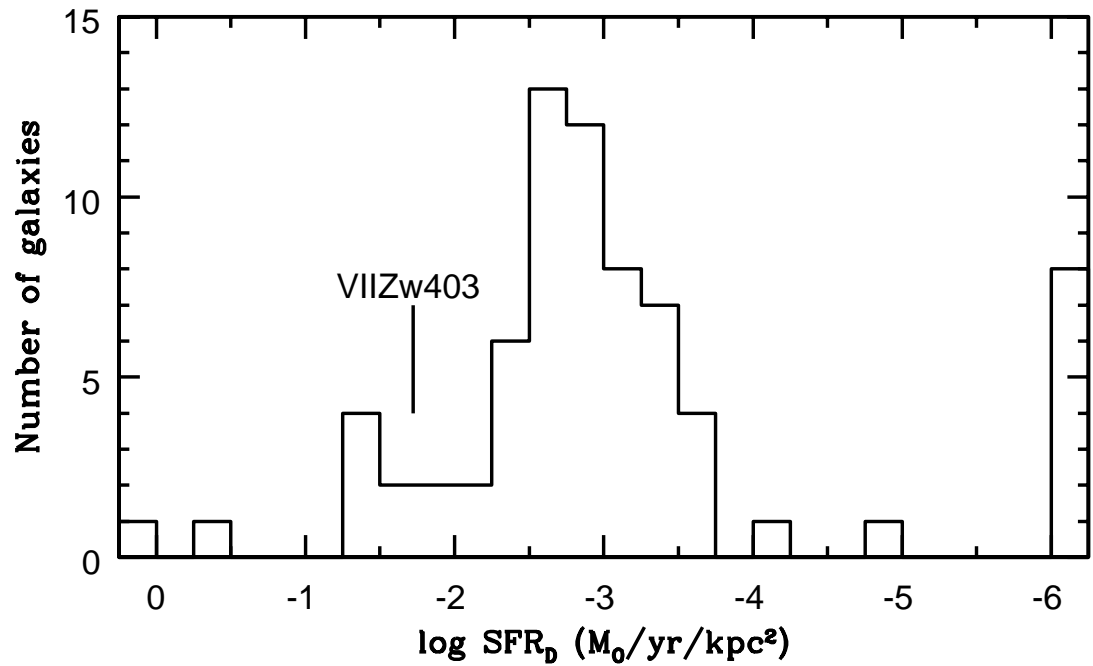}
\caption{Star formation rate per unit area (SFR$_D$; from \protect\citet{he04}
  for unbarred dIm galaxies. After Figure 11 from \protect\citet{he06}.
  SFR$_D$ 
  is the integrated H$\alpha$-based star 
  formation rate divided by the area in one $V$-band disk scale length
  of the galaxy.  VII Zw 403 has a star formation rate that is
  significantly higher than the median in this sample.
\label{fig:sfrd}
}
\end{figure}

The integrated \ha\ and UV luminosities and derived star formation
rates of the galaxy are given in Table \ref{tab:global}.  The star
formation rates derived from L$_{H\alpha}$ and the FUV luminosity use
the formulae of \citet{hunter10} that integrates from 0.1 
M\protect\solar\ to 100 M\protect\solar\ with a \citet{salpeter55}
stellar initial mass function.  The star formation rate derived from 
\ha\ is  similar to that derived from the UV.  VII Zw 403 has a
star formation rate per unit area that is normal, but towards the high
end of the range, for irregular galaxies \citep{he06}, as shown in
Figure \ref{fig:sfrd}.  Again, this is consistent with the galaxy's
classification as a BCD.  However, modeling of the color-magnitude
diagram of the stellar population by \citet{lynds98} suggests that the
star formation rate was much higher about 600--800 Myr ago than it is
today, by as much as a factor of 30.

The $V$ and 3.6 $\mu$m-band radial profiles in Figure~\ref{fig:sb} are
smooth exponential disks with about the same slope in the central part
of the galaxy, but the \ha\ and NUV are peaked in the center and fall
faster than $V$ or 3.6 $\mu$m. This indicates a change in the
primary site of star formation about 200 Myr ago, the timescale the
NUV is sensitive to. This is a relatively short dynamical time
($\approx$ two galactic rotations; see \S~\ref{sec:kinematics}) for
this slowly rotating galaxy. We cannot determine the cause of this
shift in star formation location, but it is unlikely to have been
caused by a collision or accretion event. The timescale is short
enough that we should still see either two galactic nuclei or tidal
debris, which we don't. A central collapse to a bar could explain it,
but we see no hint of a bar in the $V$, $J$, or 3.6 $\mu$m images.

At its current rate of consumption, the galaxy can turn gas into stars
for another 5 Gyr if all of the gas associated with the galaxy can be
used. The timescale to run out of gas becomes 10 Gyr if recycling of
gas from dying stars is also considered \citep{brinchmann04}.  These
timescales to exhaust the gas are somewhat short for irregular galaxies
\citep{he04}, but nevertheless, the galaxy is in no danger of running
out of the fuel for star formation any time soon.

\section{\HI\ Results} \label{sec:radio}

As \HI\ is the raw material from which molecular clouds, and then stars,
form, the \HI\ properties of a galaxy are fundamental to understanding
its past, current, and future star formation. The total \HI\ mass from
the integrated D configuration data is $5.23 
\times 10^7$ \msun, which is below the median of $3.0 \times 10^8$ for
the large sample of star-forming galaxies studied by
\citet{salzer02}. This value for the hydrogen mass 
includes only the emission we believe is associated with \vii\
itself. This does not include the secondary emission peak seen in the
spectrum around $-70$ \kms\ that we believe is due to Galactic
contamination (see \S~\ref{sec:calibration}). 

 The $M_{HI}/L_{\rm{B}}$ is 0.91. This is within the range (0.2--2.0) of
 the sample of 11 BCDs from \citet{vanzee98,vanzee01}, but is on the
 high end of the distribution for the 24 BCDs ($\sim 0.005$ to $\sim
 1.8$), and below the median for Ims ($\sim 1.3$) shown in Figure 4 of
 \citet{he04}. For comparison with the sample in \citet{salzer02}, we
 follow their method to ``correct'' the global luminosity by removing
 the effect from the starburst so that it is more representative of the
 bulk of the stars in the galaxy. This increases the $M_{HI}/L_{\rm{B}}$
 ratio to 1.82, which is higher than the \citet{salzer02} median of
 1.3. Although the \HI\ mass for \vii\ is relatively low on an absolute
 scale, its relatively high $M_{HI}/L_{\rm{B}}$ (even ``uncorrected''
 for the starburst) could indicate that it is either a somewhat gas-rich
 BCD or is currently ``luminosity-poor'' as it is presumably
 post-starburst: it should have been brighter during the burst of
 600--800 Myr ago detected by \citet{lynds98}.

\subsection{\HI\ Surface Density}

\begin{figure}
\epsscale{1.0}
\plotone{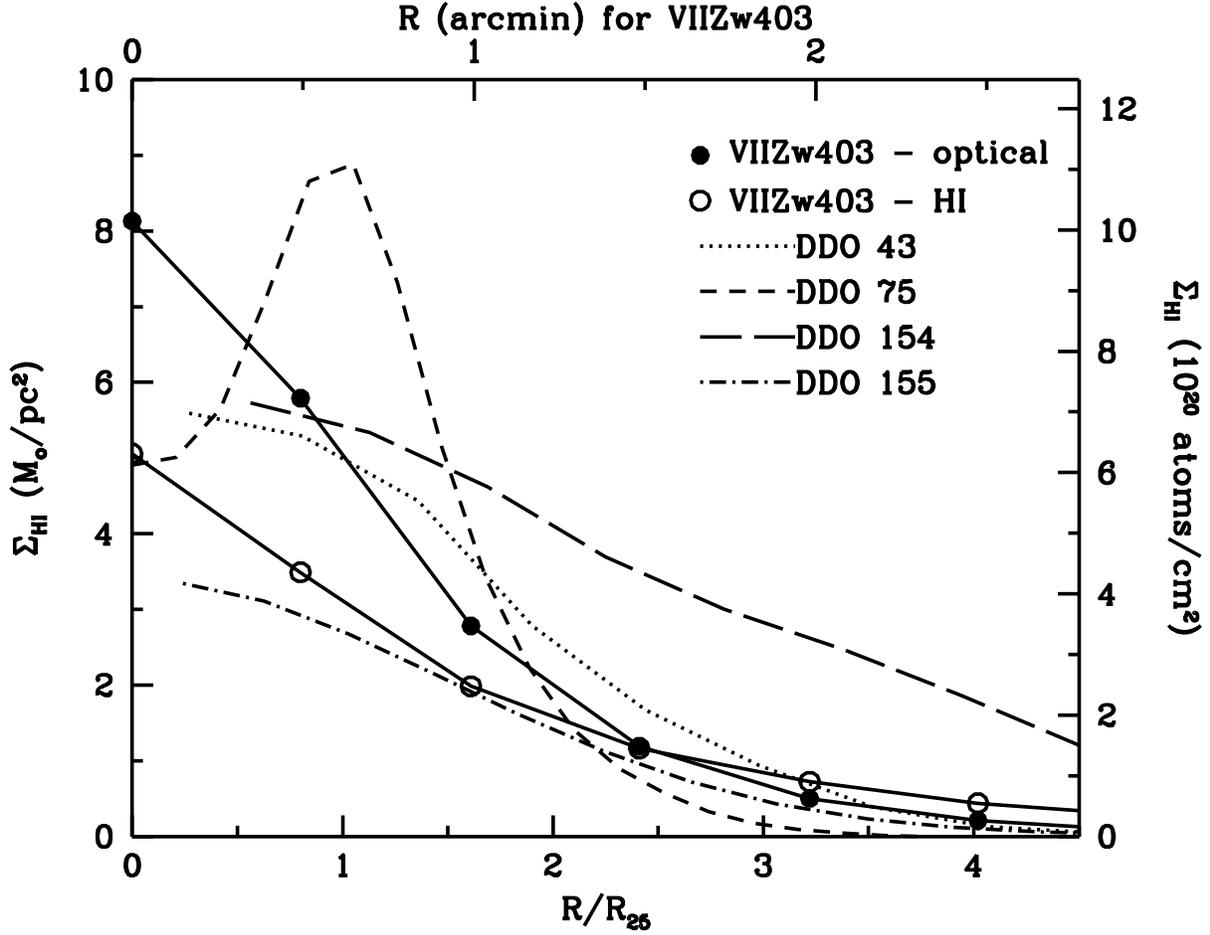}

\caption{\protect\HI\ surface density in \protect\vii\ from the
  \protect\lres\ data. The line with the solid circles labelled
  ``optical'' used the center, inclination (66\protect\arcdeg), and
  position angle ($-11$\protect\arcdeg) from the V-band data. The line
  with open circles labelled ``\protect\HI"\ used the center,
  inclination (77\protect\arcdeg), and average position angle
  (47\protect\arcdeg) from the \protect\HI\ rotation curve for the NE
  half of the galaxy. \protect\HI\ surface density profiles for galaxies
  with similar beam/$R_{1/2}$ or beam/$R_{25}$ ratios are also
  shown. Data for these profiles are from DDO 43
  \protect\citep{simpson05b}; DDO 75 (Sextans A) 
  \protect\citep{wilcots02}; DDO 154 \protect\citep{carignan89}; and
  DDO 155 \protect\citep{carignan90}.
\label{fig:surdens}
}
\end{figure}

The azimuthally averaged surface density profiles of the \HI\ gas for
\vii\ using the center, position angle, and inclination from both the
optical and \HI\ data are shown in Figure~\ref{fig:surdens}. The
profiles were calculated by integrating the \HI\ from the \lres\ map
(beam size $34\farcs1 \times 27\farcs6$) using 30\arcsec\ rings with (1)
a center position of $11^{\rm h} 24^{\rm s} 33\fs5$, 79\arcdeg\
16\arcmin\ 10\farcs3 (B1950.0), a position angle of $-11$\arcdeg, and an
inclination of 66\arcdeg\ taken from the $V$-band photometry
(\S~\ref{sec:optical}); and (2) a center position of $11^{\rm h} 24^{\rm
s}$ $35\fs1\pm10\fs3$, 79\arcdeg\ 16\arcmin\ $18\pm 27$\arcsec
(B1950.0), a position angle of $47$\arcdeg, and an inclination of
77\arcdeg\ taken from the NE half of the \HI\ rotation curve analysis
(\S \ref{sec:kinematics}).
 
Several Im galaxies are also plotted in Figure~\ref{fig:surdens}. The
galaxies were chosen to have similar linear resolutions in that the
ratio of the beam size to either the disk scale length ($R_{\rm D}$) or
the half-light radius ($R_{1/2}$) is approximately the same. The ratios
range from 1.1 to 2.4 for beam/$R_{\rm D}$ and 0.5 to 1 for
beam/$R_{1/2}$. The values for \vii\ are 1.4 and 1.2, respectively.  We
can see from the figure that the \HI\ surface density for \vii\ derived
using the \HI\ parameters falls
off smoothly and slightly more steeply in the central regions than those
for the Ims; more so when using the optical parameters. In comparison to
the BCD and dIrr \HI\ surface density 
profiles presented in \citet{vanzee98}, the shape of \vii's profile is
similar to those of the BCDs, and somewhat steeper than those for 
the dIrrs. The central surface densities from both the optical and \HI\
parameters for \vii\ are seen to be low compared to the BCDs in the
sample; when using the \HI\ parameters, the central surface density is
like that of the dIrrs. 

\subsection{Morphology} \label{sec:HImorph}

\begin{figure}
\includegraphics[scale=0.70,angle=-90]{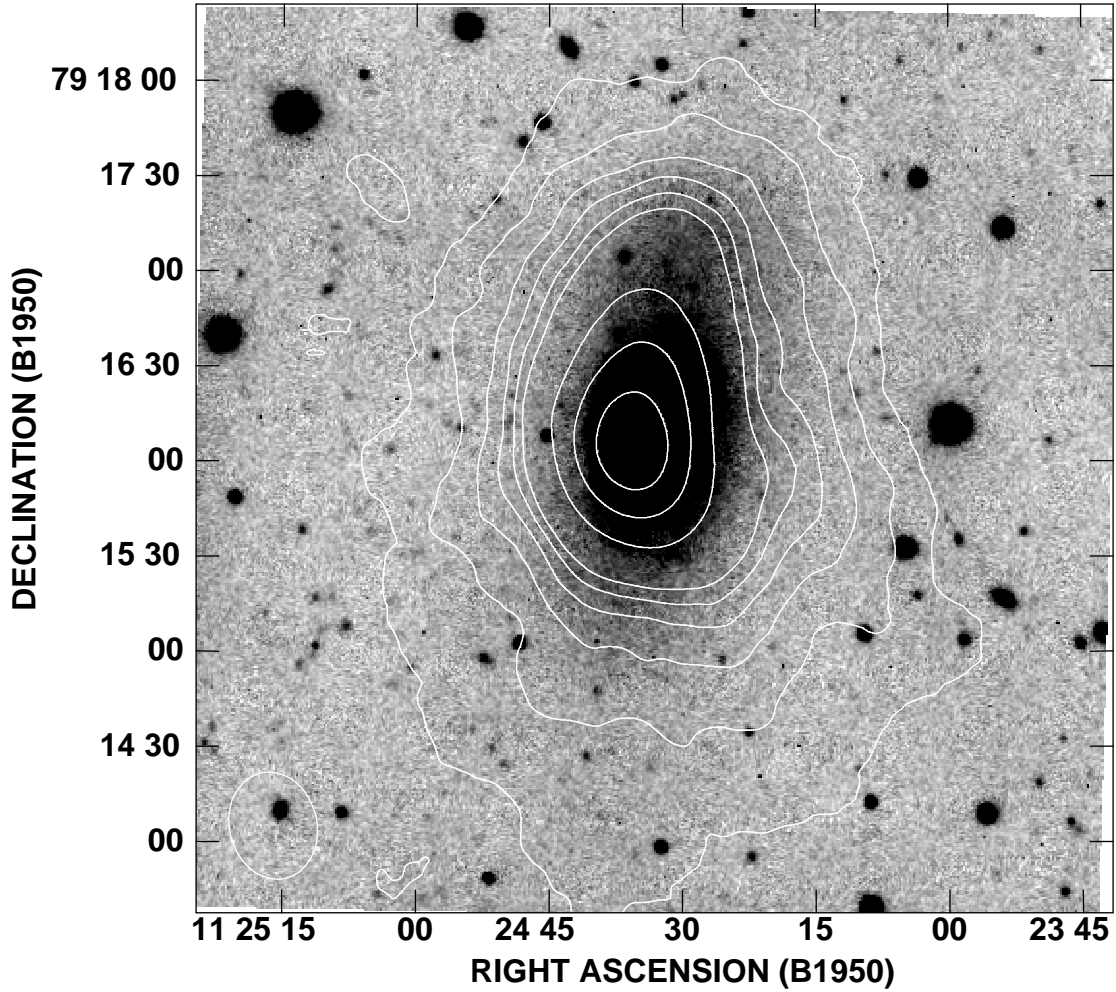}
\caption{The V image overlaid with contours from the \protect\lres\
  configuration integrated flux map. The beam size is indicated in the
  lower left corner. The levels are (2, 10, 20, 30, 40, 50, 100, 150,
  and 200) $\times 10^{19}$ \protect\acm. The lowest contour is at
  1.3$\sigma$. (A single-channel 2\protect$\sigma$\ detection integrated
  over 3 channels is equivalent to a column density of 1.5 $\times
  10^{19}$ \acm, which we take as indicative of the noise level in the
  integrated map.) The field of view is approximately 5\protect\arcmin
  $\times$ 5\protect\arcmin.
\label{fig:cd1m0onv}
}
\end{figure}

Figure~\ref{fig:cd1m0onv} shows the integrated \lres\ \HI\ contours on
the V image, with the contrast set to show the full extent of the
galaxy's optical emission. This figure shows that there is a large,
low-level irregular skirt of \HI\ gas that extends beyond the optical
emission, especially toward the south. Irregular gas distributions that
extend beyond the optical emission, which are common in field galaxies,
are also found in other BCDs
\citep{taylor94,vanzee98,vanzee01,thuan04}. From the D configuration
data (not shown), the total extent (diameter) of the \HI\ ($2 R_{HI}$)
out to the $1 \times 10^{19}$ \acm\ level is 5\farcm24
(deconvolved). This is 2.3 $R_{\rm{H}}$, which is typical for the Ims in
Figure 13 of \citet{hunter97}.

At the spatial resolution of the \lres\ image ($\sim 700 \times 600$
pc), the \HI\ appears smoothly distributed with the density increasing
toward the center of the galaxy.  This smoothly increasing density is often
seen in BCDs \citep{taylor94,simpson00,vanzee98,vanzee01}; and although
some ``clumpiness'' of the \HI\ is also seen in BCDs \citep{vanzee98},
it is not like the highly-structured gas distribution often observed in
dIrrs, where shells, filaments, and loops are frequently visible
\citep{meurer92,puche92,young96,martin98,meurer98,wb99,stewart00,ott01,
simpson05a,simpson05b}. Comparison of the \HI\ distribution in \vii\ to
those in dIrrs with similar linear spatial resolution, such as
DDO 88 \citep{simpson05a} and DDO 43 \citep{simpson05b}, shows that at
these scales, this increased ``smoothness'' in \vii\ is real. Clumping
of the \HI\ on the scales seen in dIrrs is not visible in \vii.

\begin{figure}
\epsscale{0.9}
\plotone{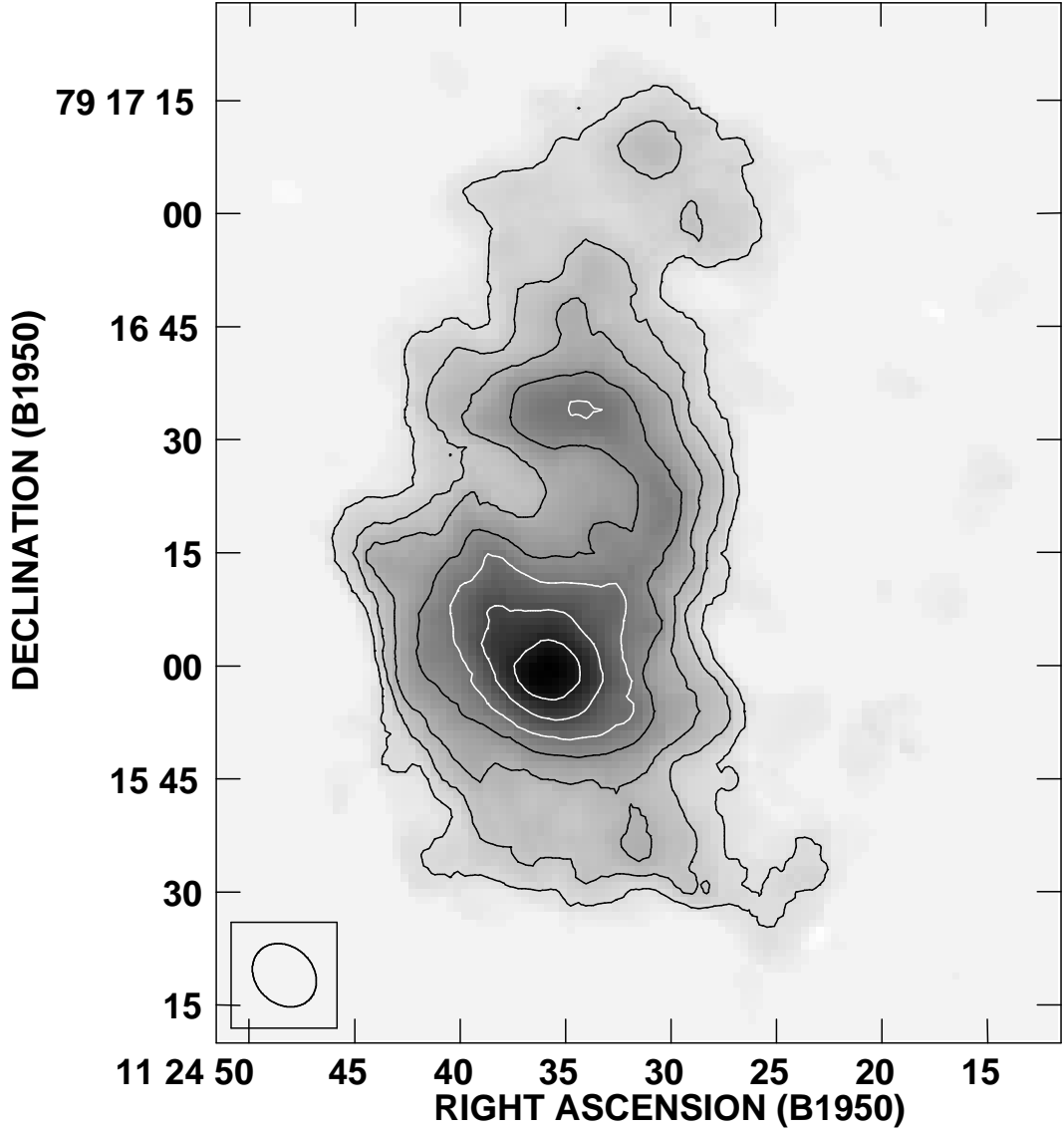}
\caption{Integrated flux density map from the \protect\mres\
  configuration data. The beam size is indicated in the lower left
  corner. The contour levels are (5, 10, 15, 23, 30, 40, 50) $\times
  10^{20}$ \protect\acm.  The lowest contour is at 2.4$\sigma$. (A
  single-channel 2\protect$\sigma$\ detection integrated over 3 channels
  is equivalent to a column density of 2.12 $\times 10^{20}$ \acm, which
  we take as indicative of the noise level in the integrated map.) The
  field of view is 1.9\protect\arcmin\ $\times$\ 2.3\protect\arcmin,
  which corresponds to 2.4 $\times$\ 2.9 kpc.
\label{fig:bcd0m0}
}
\end{figure}

At higher spatial resolution, however, some structure begins to be
revealed in the \HI. Figure~\ref{fig:bcd0m0} shows the integrated \HI\
flux from the \mres\ data cube, with a linear resolution of 200 $\times
160$ pc ($9\arcsec \times 8\arcsec$), as compared to the
750 $\times$\ 600 pc resolution of the \lres\ image. (Please note the
change in field of view of the two images; the \mres\ image shows only
the inner $2\arcmin$, which corresponds roughly to the area covered by
the central three contours in the \lres\ map shown in
Figure~\ref{fig:cd1m0onv}.) This image shows that the central \HI\
distribution contains a region of lower density separating a large clump
to the south of the center and a much smaller clump to the north,
connected by a ridge on the west side. This low density region resembles
a partial cavity or hole, with decreasing density to the east side. This
is further discussed below. 

\subsection{The Velocity Field} \label{sec:kinematics}

\begin{figure}
\includegraphics[scale=.65,angle=-90]{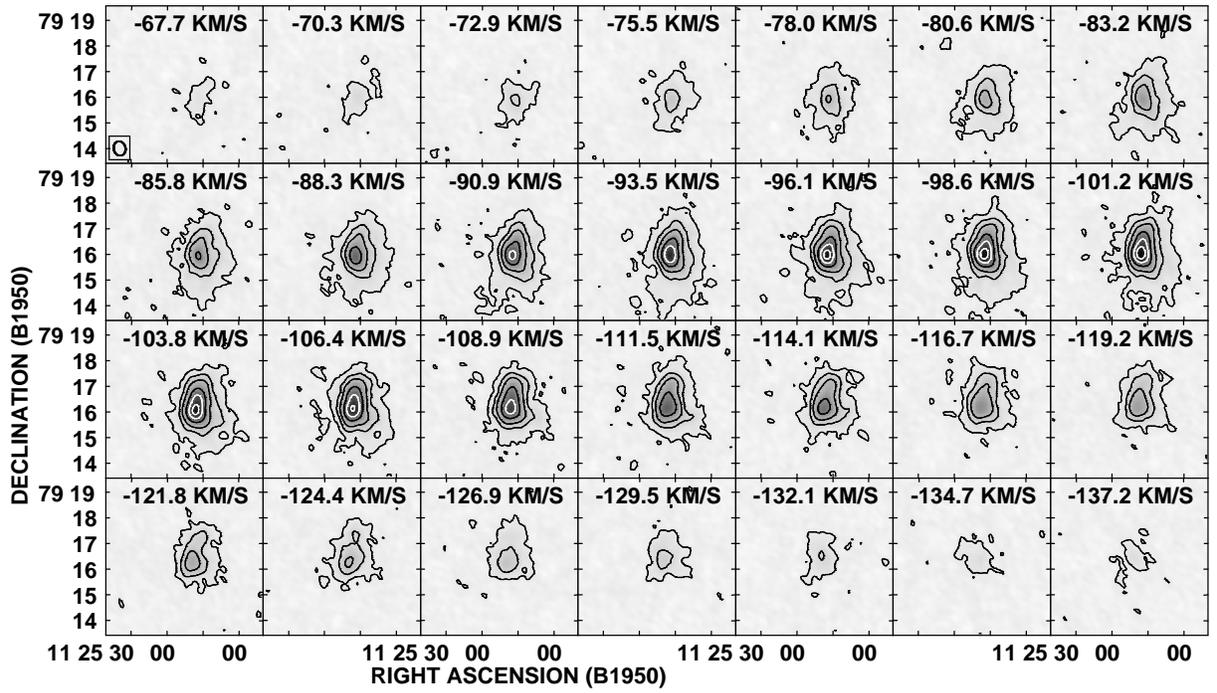}
\caption{Channel maps of the \protect\HI\ emission in the \protect\lres\
  configuration data cube. The velocity of each channel is printed along
  the top; the beam size is indicated in the first (top
  left) channel. The contours are at (3, 10, 20, 40, 60, 80, and 100) $\times$
  the single-channel r.m.s. (0.83 mJy beam$^{-1}$).
\label{fig:cd1chanmaps}
}
\end{figure}

\begin{figure}
\epsscale{1.0}
\plotone{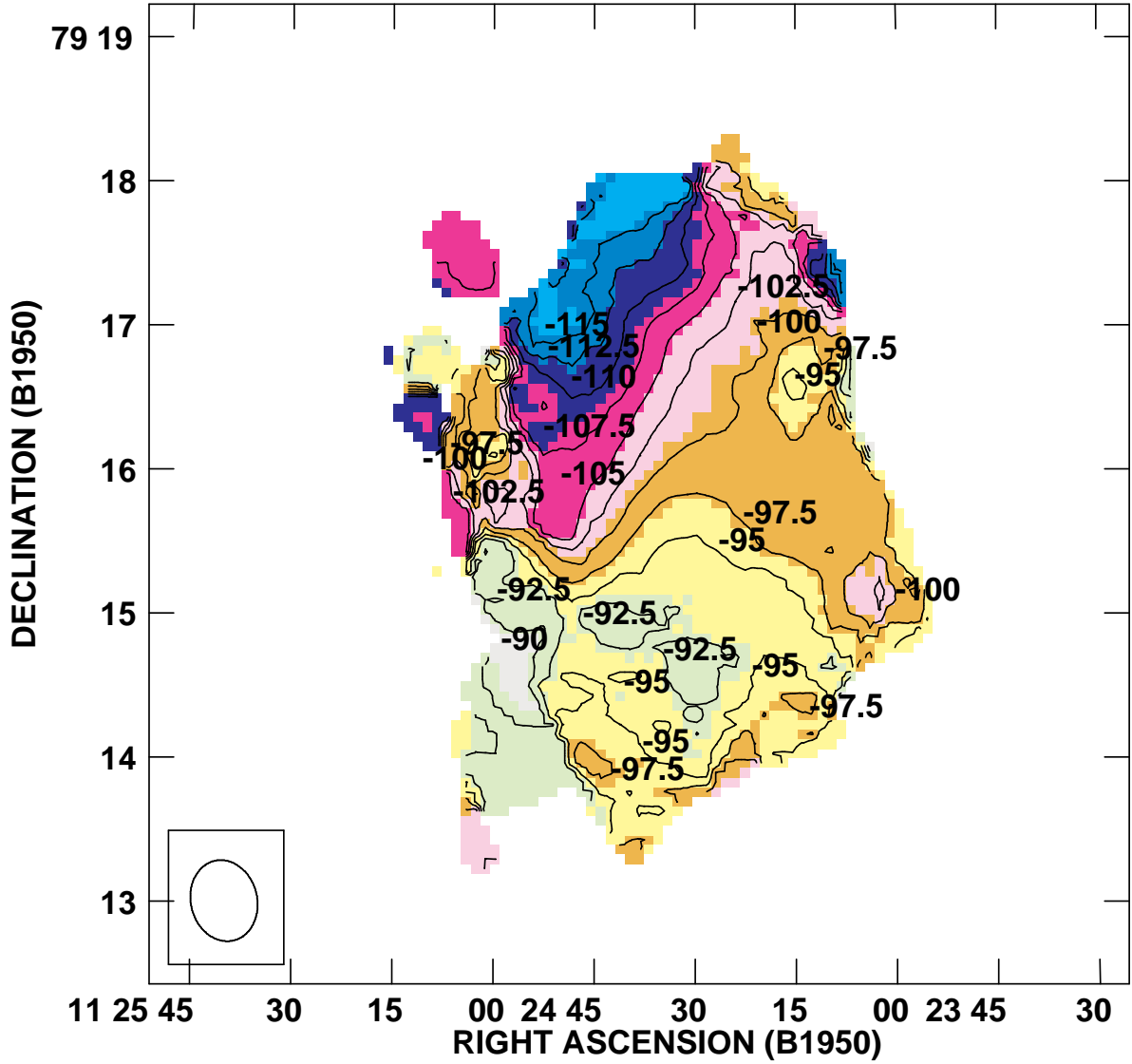}
\caption{\protect\HI\ velocity field from the \protect\lres\
  configuration data. The beam size is indicated in the lower left
  corner. Contours are in \protect\kms.
\label{fig:cdm1}
}
\end{figure}

\begin{figure}
\includegraphics[scale=0.75,angle=-90]{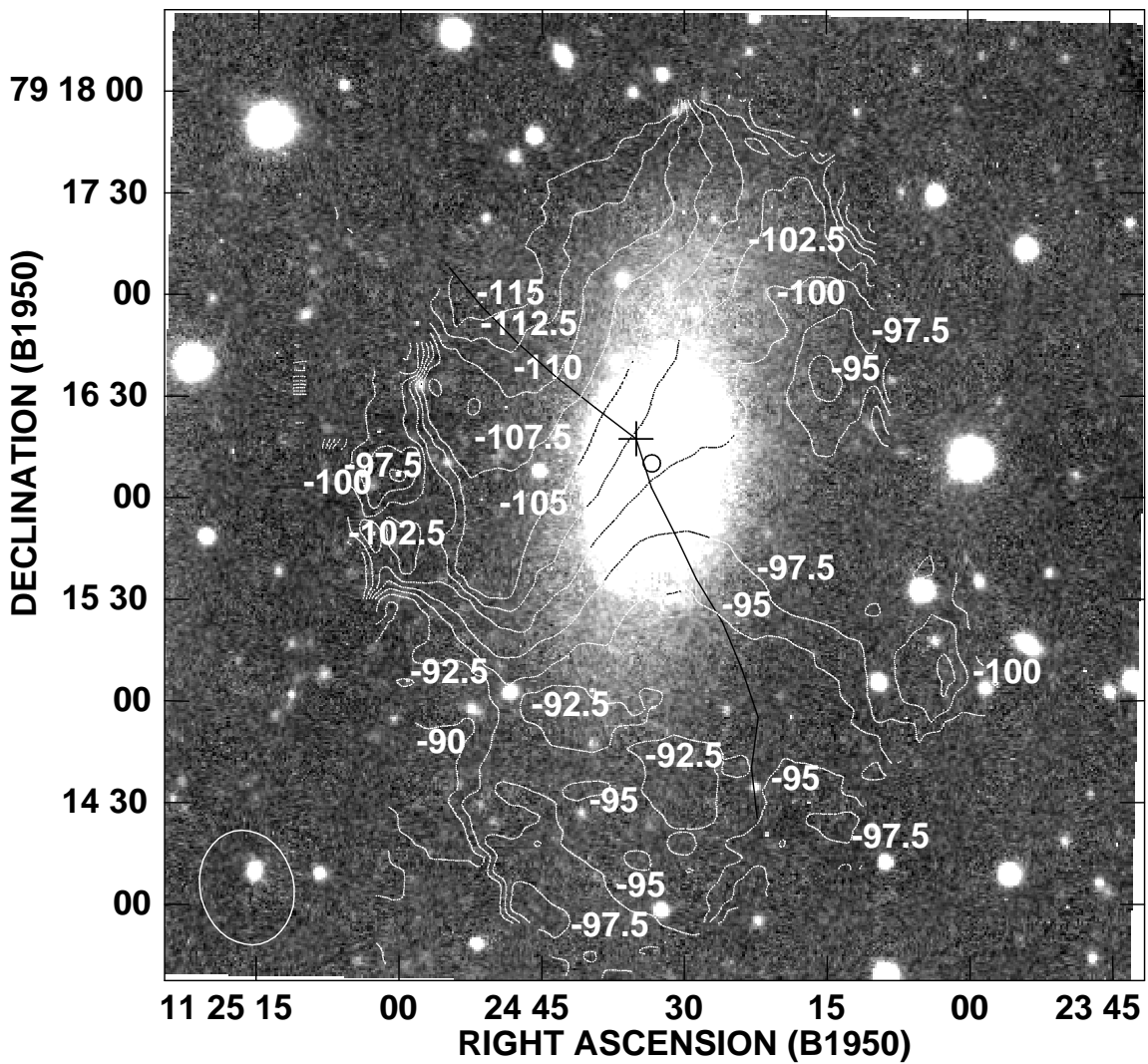}
\caption{Isovelocity contours from the \protect\lres\ configuration data
moment one map superposed on a V-band image of VII Zw 403. The contours,
in white, are labeled with the velocity in km s$^{-1}$.  The black line
running more or less perpendicular to the isovelocity contours shows the
position 
angle of the rotation axis in 15\protect\arcsec\ steps.  The plus sign
marks the kinematic center of the galaxy, and the open circle marks the
center in the V-band.  Note that the uncertainties on the kinematic
center ($\approx 27$\arcsec) are approximately the same size as teh FWHM
of the beam, which is indicated by the ellipse in the lower left corner
(34.1\protect\arcsec$\times$27.6\protect\arcsec).
The velocity resolution is 2.6 km s$^{-1}$.  
\label{fig:velfield}
}
\end{figure}

In Figure~\ref{fig:cd1chanmaps} we show the channel maps from the \lres\
data cube that were integrated to produce the flux-weighted \HI\
velocity field shown in Figure \ref{fig:cdm1}, and in Figure
\ref{fig:velfield} as isovelocity contours superposed on the V-band
image of VII Zw 403. One can see here that there is ordered rotation,
with the NE portion of the galaxy approaching and the SW half of the
galaxy receding. However, it is also immediately apparent that the
velocity field is complex.  First, there is a curve in the
isovelocity contours so that the NE and SW portions of the galaxy have
distinctly different kinematic axes. The kinematic axis of the NE half
of the galaxy is aligned roughly at an angle of +45\arcdeg. The SW part
of the galaxy, on the other hand, appears to have a line of nodes at
about +25\arcdeg, which is aligned more closely with the optical galaxy's
major axis at $-11$\arcdeg.  Second, there are places where the velocity
field is not well behaved. In particular on the east side of the galaxy
the isovelocity contours curve back upward in a complex manner, and in
the south there are ``islands'' of different velocities.

As strange as this velocity field appears, there are other dwarf
galaxies with curves in the position angles of their isovelocity
contours similar to what we see in VII Zw 403.  A review of the literature
revealed ten other BCDs with this type of feature
(ESO338-IG04: \citealt{ostlin01};
Haro 2: \citealt{bravo04};
Haro 33, Mrk 51: \citealt{simpson00};
NGC 1569: \citealt{stil02b};
NGC 3738: \citealt{stil02a};
NGC 1705: \citealt{meurer98};
UGC 4483, UM462: \citealt{vanzee98};
UM323: \citealt{vanzee01}),
and there are undoubtedly others. 
However, this phenomenon is not exclusively that of BCDs.
We also found a comparable number of Im galaxies with this
sort of feature
(DDO 22, DDO 68: \citealt{stil02a};
DDO 69: \citealt{young96};
DDO 165, DDO 187, UGC 3698: \citealt{swaters99};
DDO 210: \citealt{begum04};
DDO 216: \citealt{young03};
ESO 245-G005: \citealt{cote00};
NGC 2366: \citealt{hew01};
NGC 4163: \citealt{simpson00}).
Since some of the Im galaxies have no current star formation activity or
very little, we cannot associate this velocity field feature with the
starburst that makes VII Zw 403 special.  In addition, this is not
relegated only to slow rotators, such as VII Zw 403. In the sample taken
from the literature, maximum rotation speeds range from very small up to
69 \kms.  Distortions in velocity fields are often blamed on
gravitational interactions with other galaxies. In these galaxies, some
are known to have small companions and others are not.

To get some idea of the rotation of VII Zw 403, using the task {\sc gal}
in {\sc aips} we fit the two halves of
the velocity field separately and began with the NE or approaching half.
We used the \lres\ moment map which has a beam size of
34\farcs1$\times$27\farcs6.  With the \HI\ peak as a first guess on
the position of the kinematic center, we fit the field with a Brandt
function \citep{brandt60}. We also experimented with changing the
initial conditions: we varied the position angle from 34--65\arcdeg,
the inclination from 66\arcdeg to 72\arcdeg, and maximum rotational
velocity from 
13 tp 15 \kms. We also varied the radial extent of the fit to raidd of
130\arcsec, 90\arcsec, and 75\arcsec; the center up to eight pixels; and
the systemic velocity by a few \kms. We tried fitting the whole field
and just one side at a time while varying initial parameters. For the
final rotation curve(s) we used a position angle of 40\arcdeg, an
inclination of 66\arcdeg, and a rotational velocity of 13 \kms\ for the
initial guesses. From this we determined the center (11$^{\rm h}$
24$^{\rm m}$
35\fs1$\pm10\fs3$, 79\arcdeg\ 16\arcmin\ 18$\pm$27\arcsec, B1950.0) and
systemic velocity ($-103\pm1$ \kms).  The experiments with varying
initial guesses suggested that the uncertainties given above are
reasonable.  A fit with the center fixed and a solid body function
yielded the same systemic velocity.  We then fixed the center position
and central velocity and fit the field in concentric rings that were
15\arcsec\ in width and stepped every 15\arcsec.  The innermost circle
did not yield a solution and that point is omitted.  For the SW or
receding half of the galaxy, we adopted the center position and systemic
velocity determined from the NE half and again fit in concentric rings.

\begin{figure}
\epsscale{0.75}
\plotone{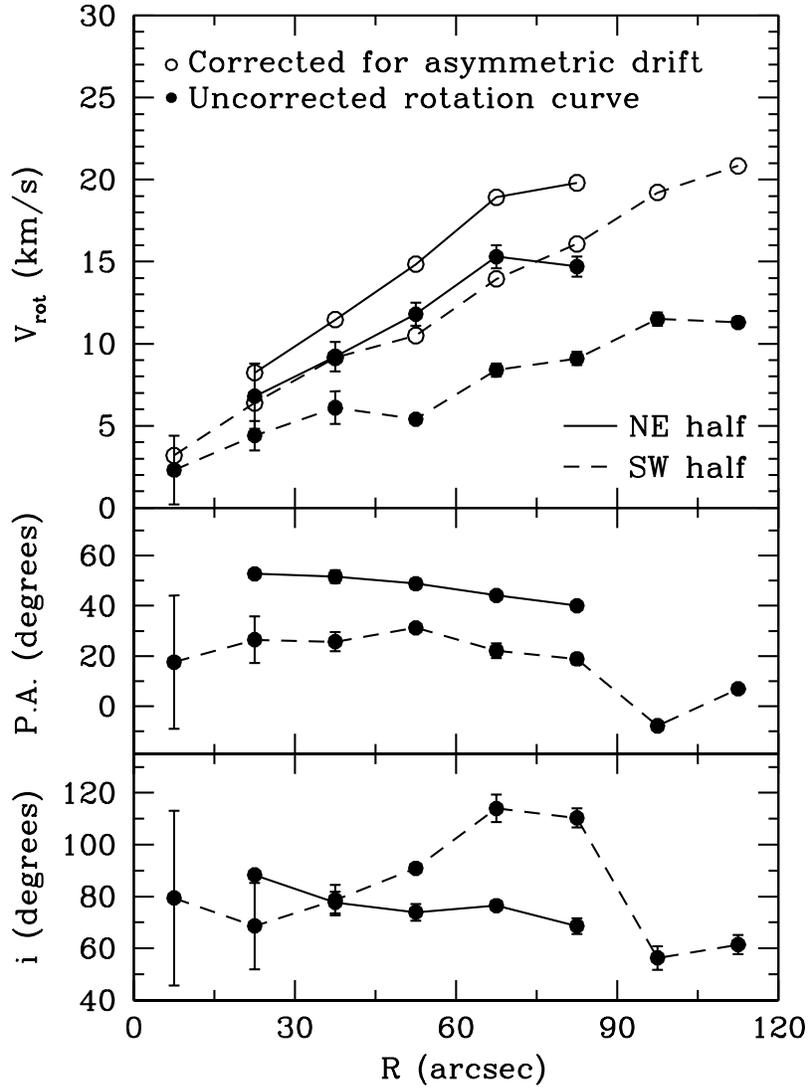}
\caption{Best-fit rotation curve for VII Zw 403. The velocity field was
fit in two separate parts: the NE or approaching half and the SW or
receding half. Only the central position and systemic velocity were
fixed in these fits, and the rotation speed, position angle, and
inclination were determined in 15\protect\arcsec\ wide annuli stepped in
15\protect\arcsec\ intervals from the center. The solid  symbols
are the original rotation curve; the open symbols show the curve
corrected for pressure support (asymmetric drift correction).
\label{fig:rot}
}
\end{figure}

The results are shown in Figure \ref{fig:rot} as solid circles.  The
kinematic axis in 
each ring is traced as the black line on Figure \ref{fig:velfield}.  One
can see that for each half alone the position angle is relatively stable
with radius except for the outer two annuli on the SW side.  The
inclination, on the other hand, is quite variable although over much of
each half of the field it changes systematically. Generally the
inclinations are all higher than we expect based on the optical
major-to-minor axis ratio. The maximum rotation speed is about 15 \kms\
on the NE side and a few \kms\ lower on the SW side. 

\begin{figure}
\epsscale{1.0}
\plotone{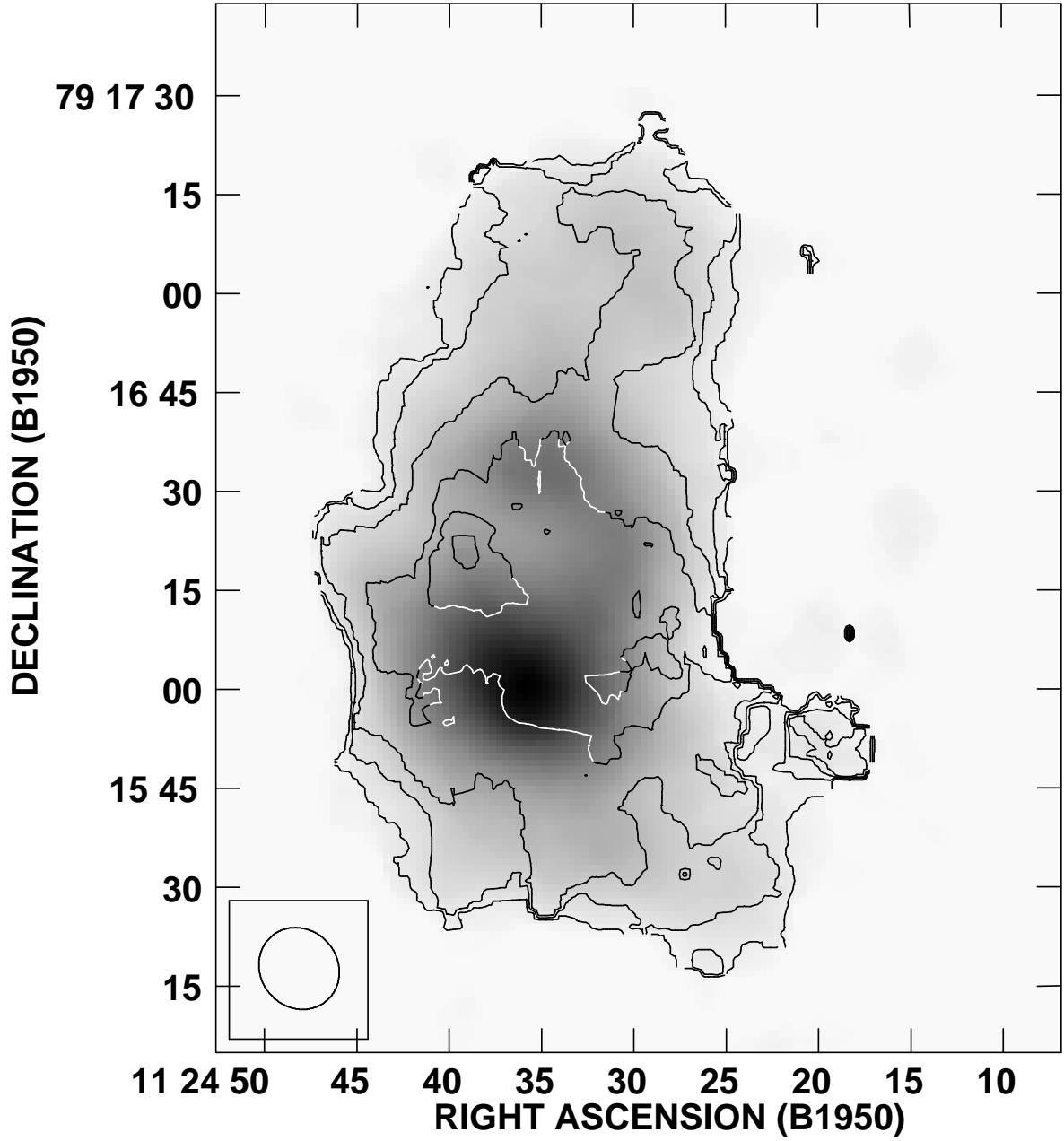}
\caption{Velocity dispersion contours superposed on the integrated flux
   from \protect\bcdone\ data. The beam size is indicated in the lower
   left corner. Contours are 4, 6, 8, 10, 12, and 14 \protect\kms.
\label{fig:veldisp}
}
\end{figure}

With a maximum rotation velocity of only $\sim$15 \kms\ ($\sim$ 16 \kms\
when corrected for an inclination $i = 66\arcdeg$) and an average
dispersion velocity in the \lres\ map of 8.5 \kms\
(Figure~\ref{fig:veldisp}; \S \ref{sec:disp}), the galaxy has a 
significant amount of pressure support from the random motions in the
gas. This means that simply using the rotation velocity will result in
an underestimate of the dynamical mass. We have corrected for this by
adding in the pressure support; this is called the asymmetric drift
correction.  To calculate this, we used the procedure described in
\citet{begum04} and we follow their discussion here. To make this
applicable to a gas disk rather than a collisionless stellar system for
which the rotational velocities are much larger than the random motions,
we make 
the assumption that the pressure support can be approximated by the gas
density times the square of the velocity dispersion:
\begin{equation}
  v_{c}^{2} = v_{0}^{2} - r 
  \sigma^{2}[\frac{\rm d}{{\rm d}r}(\ln\Sigma_{\rm HI})
  + \frac{\rm d}{{\rm d}r}(\ln\sigma^{2}) - \frac{\rm d}{{\rm d}r}(\ln2h_{z})].
\end{equation} 

Here, $v_c$ is the corrected circular velocity, $v_0$ is the observed
rotation velocity from the original rotation curve, $\Sigma_{\rm HI}$ is
the \HI\ surface density, $\sigma$ is the velocity dispersion, and $h_z$
is the scale height of the \HI\ disk. We further assume that neither the
velocity dispersions nor the scale height vary significantly across the
disk (i.e.\ with radius $r$), so their derivatives are zero, leaving us
with
\begin{equation}
v_{c}^{2} = v_{0}^{2} - r \sigma^{2}[\frac{\rm d}{{\rm
      d}r}(\ln\Sigma_{\rm HI})]. 
\label{asym}
\end{equation}

To determine the derivative of the \HI\ surface density, we must
determine a functional expression to describe it. Towards this end, we
fit the profile (derived using the optical center) shown in Figure \ref{fig:surdens} with a Gaussian of the
form: 
\begin{equation}
\Sigma_{\rm HI}(r) = \Sigma_{0}\, {\rm e}^{{-(r-\mu)^2}/{2r_{0}^2}}
\end{equation}
where $\Sigma_0$ is the amplitude, $\mu$ is the radius at which the
Gaussian peaks, and $r_0$ is the
scale length. The profile was fairly well fit with $\Sigma_0 =
11.42\pm0.54$ \msun\ pc$^{-2}$, $\mu = 26\pm6\arcsec$ (0.55 kpc),
and $r_0 = 57\pm3\arcsec$ (1.26 kpc). 
Taking the derivative, eq. \ref{asym} now becomes
\begin{equation}
v_{c}^{2} = v_{0}^{2} + \frac{r(r-\mu)\sigma^2}{r_{0}^2}.
\end{equation}
Using the values found from the gaussian fit and a galaxy-wide average
of 8.0 \kms\ as a conservative lower estimate for the velocity
dispersion (corrected for observational broadening due to the velocity
resolution), we then corrected the rotation curves for both the NE and
SW halves. The result is shown in Figure \ref{fig:rot} (open circles),
along with the uncorrected rotation curves (solid circles). Applying the
correction increases the maximum velocities to 20 \kms\ for the NE half
and 22 \kms\ for the SW half. The corrected rotation curve is
essentially that of a solid body on both sides of the galaxy.

The necessity of correcting the rotation to include pressure support is
apparent when we calculate a dynamical mass for \vii. Using the highest
rotational velocity from the uncorrected rotation curves and $M = V_{\rm
max}^2 R_{\rm max}/G$ we get an estimated dynamical mass of $7.9 \times
10^7$ \msun. This is an $M_{\rm HI}/M_{\rm dyn}$ of 0.66, which is higher
than the ranges for BCDs found by others
\citep{taylor94,vanzee98,vanzee01}. The problem becomes worse when the
total luminous mass ($M_{\rm gas} + M_{\rm stars}$) is considered. The
gas mass is found by multiplying the \HI\ mass by a factor of 1.34 to
correct for He. No correction is made for molecular gas; no good
estimates exist as CO detections
in BCDs have been few, and those primarily in BCDs with elliptical inner
isophotes \citep{sage92}, or in metal-rich dIrrs: \citet{taylor98} report
from a literature sample that there are no known detections of CO in
galaxies with $12 + \log(O/H) < 7.9$. \vii\ has irregular inner isophotes
and $12 + \log(O/H) = 7.69$ \citep{izotov99}, so is unlikely to have
detectable CO.

To find the mass in stars, we can estimate a lower limit by using a
stellar $M/L_V$ ratio from \citet{bell01} that is a function of
$(B-V)_0$ color: $\log(M/L_V) = -0.37 + 1.14 (B-V)_0.$ This is
appropriate for a \citet{salpeter55} stellar initial mass function and
metallicity $Z=0.008$ and is derived using the \citet{bruzual03} stellar
population models.  This results in a stellar mass of approximately $3.9
\times 10^7$ \msun, or a stellar mass-to-light ratio of 0.89. However,
this does not take the starburst into account. Assuming that the SFR was
30 times higher for 200 Myr \citep{lynds98} during the past 13 Gyr, we
then add that to the constant SFR model (0.013 \msun\ yr$^{-1}$) and obtain
an upper estimate of $9 \times 10^7$ \msun\ (stellar $M/L_{\rm B}$ =
1.6). With this range, our luminous mass estimate is between $1.1 \times
10^8$ and $1.6 \times 10^8$ \msun. Even the lower of these is greater
than the highest dynamical mass estimated from the uncorrected rotation
curve by a factor of 1.4.

However, when we use the corrected rotation curves to calculate a
dynamical mass, we get $M_{\rm dyn} = 2.5 \times 10^8$ \msun, which is
1.5 times larger than even the higher estimate of the luminous mass. This
is now a $M_{\rm HI}/M_{\rm dyn}$ of 0.21, which is more in line with
those in \citet{taylor94} and \citet{vanzee98,vanzee01}.

\subsection{Velocity Dispersions}\label{sec:disp}

\begin{figure}
\epsscale{0.85}
\plotone{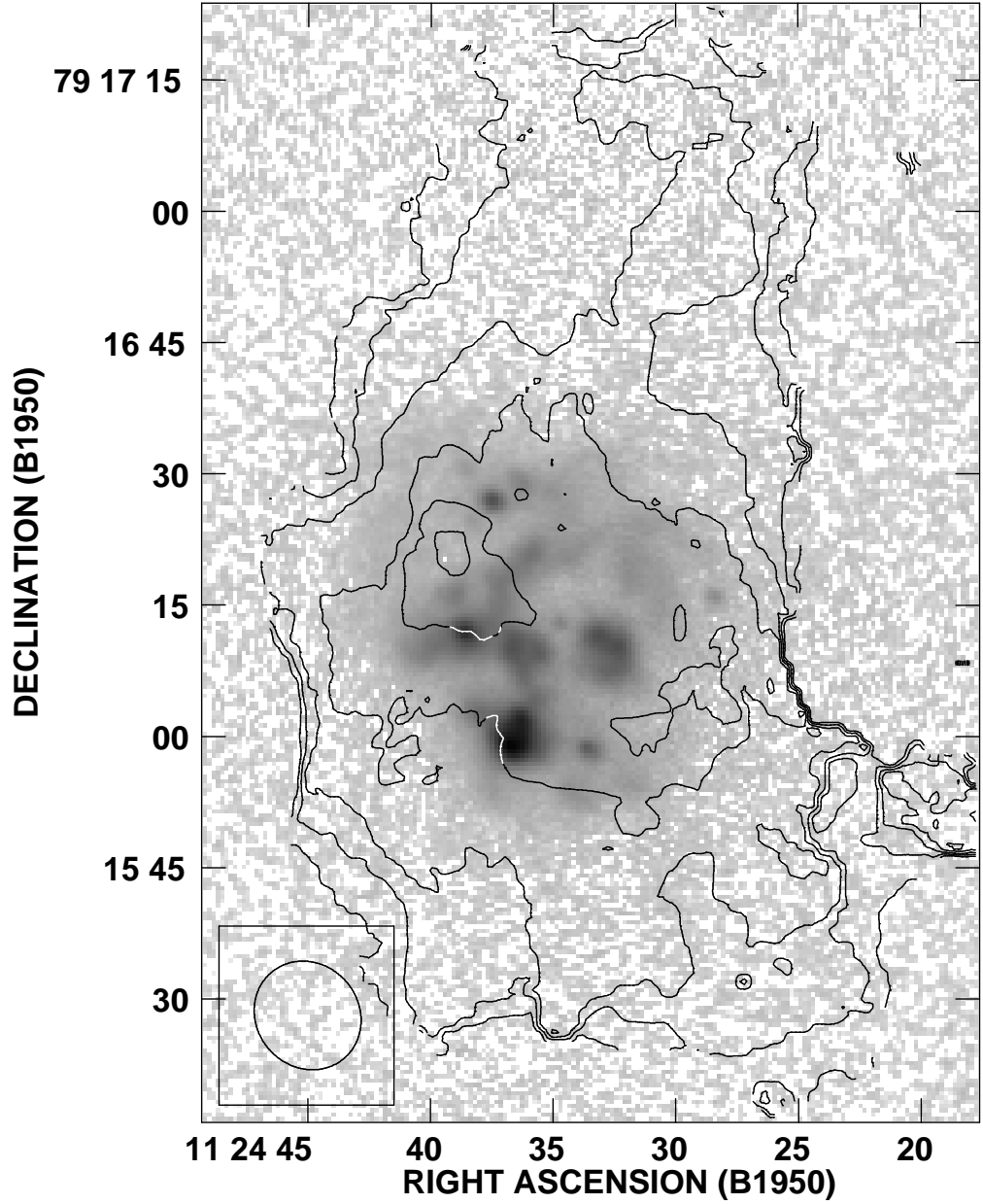}
\caption{The \ha\ image with contours from the \protect\bcdone\ velocity
  dispersion map. The beam size is indicated in the lower left
  corner. Contour levels
  are 4, 6, 8, 10, 12, and 14 \protect\kms. 
\label{fig:bcd1m2onha}
}
\end{figure}

Contours from the \bcdone\ data showing the velocity dispersions across
the galaxy overlaid on the \HI\ flux map are shown in Figure
\ref{fig:veldisp}. In general, the higher dispersion areas ($\sim 10$
\kms) correspond to the area where most of the \ha\ flux originates, as
shown in Figure \ref{fig:bcd1m2onha}. There is nothing notable in either
the $V$ or \ha\ images in the unresolved region of highest dispersion
(located just NE of the large \HI\ clump), but it is adjacent to an
unresolved x-ray source \citep{ott05a}, as discussed below.

\section{Discussion}

\subsection{Relation between Star Formation and Gas}

\begin{figure}
\plotone{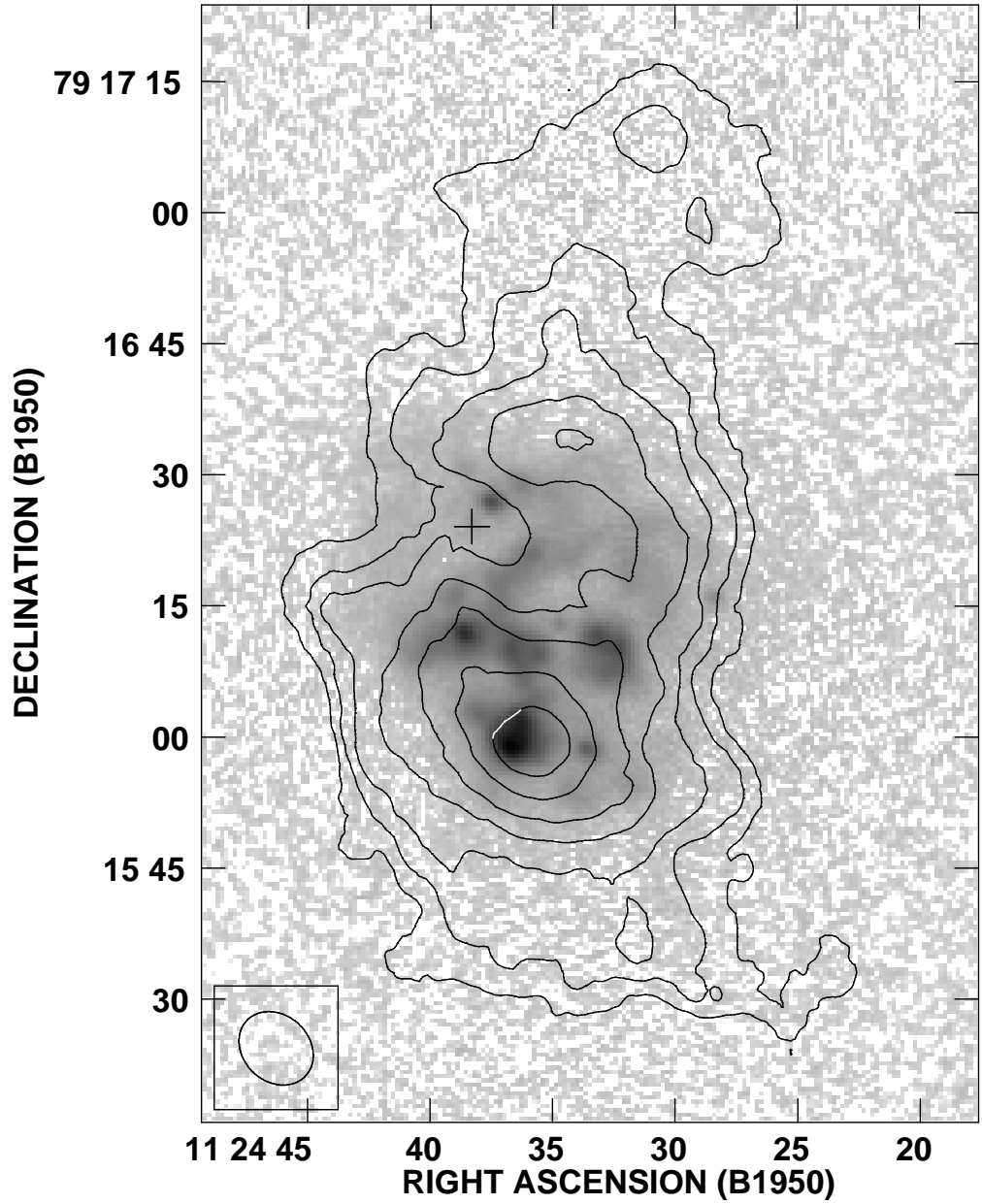}
\caption{The \protect\ha\ image overlaid with contours from the
  \protect\mres\ configuration data. The beam size is indicated in the
  lower left corner. The plus sign marks the location of the
   unresolved x-ray source detected by \protect\citet{ott05a}. The contour
  levels are the same as for 
  Figure~\ref{fig:bcd0m0}.  
\label{fig:bcd0m0onha}
}
\end{figure}

As can be seen in Figure~\ref{fig:bcd0m0onha}, the highest \HI\
column density region roughly corresponds to the region of strongest \ha\
emission. This is seen in other BCDs as well \citep{vanzee98}. There is
diffuse \ha\ emission that spreads to the north, but not as far as the
\HI\ emission. \citet{thuan04} find that there is weak
continuum emission (1.4 GHz) associated with the \ha\ emission;
with a slight extension to the north at 1.4 GHz as
well. Without a spectral index, they were unable to verify that the
continuum emission is mainly thermal in nature (and thus directly due to
to the \HII\ region), but it does seem likely.

\citet{ott05a,ott05b} presented {\it Chandra} x-ray observations of
several dwarf starburst galaxies, including \vii. Their data show no
diffuse x-ray emission associated with \vii. They found that the diffuse
x-ray emission they detected in the other galaxies in their sample was
probably associated with the development of galactic winds from coronal
gas heated by star formation.  They suggest that the lack of detected
diffuse x-ray emission for \vii\ (they give an upper limit of $ 2.8\times
10^{38}$ erg s$^{-1}$) could be due to several possible factors,
including low metallicity or smaller volume.  The expanding \ha\ shells
in the two galaxies with no detected diffuse x-ray emission (\vii\ and I
Zw 18) are 10-17 Myr older than those in the galaxies for which they
do detect diffuse emission. Perhaps a coronal wind never developed, or
perhaps it has already had time to cool.

\citet{ott05a} did detect one unresolved x-ray source whose power-law
spectral index is consistent with that of an x-ray binary, but did not
detect the filaments or lobes seen by \citet{papa94} from ROSAT
data. Similar x-ray sources in the \citet{ott05a} sample of 8
starbursting galaxies are located near ``bright \HII\ regions, rims of
superbubbles or young stellar clusters.'' The location of the unresolved
x-ray source relative to the \ha\ emission can be seen in
Figure~\ref{fig:bcd0m0onha}. The x-ray source is located near, but not
coincident with, an \ha\ knot. As seen in Figures~\ref{fig:bcd1m2onha}
and \ref{fig:bcd0m0onha}, the x-ray source is also on the north edge of
the high dispersion knot, and is located in/on the northeast side of the
depression/cavity in the \HI, in a smaller region of even lower
density. Recall that the broadband surface photometry
(Figure~\ref{fig:sb}) was consistent with constant star formation in the
outer regions over the past 10--20 Gyr, with more recent star formation
(past 200 Myr) in the central regions. 
If the x-ray source is, indeed, an x-ray 
binary, then perhaps we are seeing an evolved stellar system located
near a region of recent star formation, and also near a region of
disturbed neutral gas. The logical explanation is that a star formation
episode and/or resulting supernova has injected energy into the \HI,
increasing the dispersion; and depleting or perhaps clearing out some of
the neutral gas.

To test this, we examined the kinematics of the gas in the \mres\ cube
in a small region around the x-ray source. To enhance the
signal-to-noise in the spectra, we ``collapsed'' a six-pixel square
region centered on the x-ray source to a single pixel, and then examined the
resulting spectrum. There was a fairly clear double-peak visible, so we
fit a pair of gaussians to them (using {\sc xgaus} in
{\sc aips}). The two peaks are separated by 13.5 \kms. This is
consistent with expansion (or contraction) of $\sim 7$ \kms. The
apparent radius of the depression around the x-ray source in the \mres\
data is $\sim 105$ pc, which is roughly the same size as
the beam. This means that the region may not be resolved, so this size
is an upper limit. If we assume a constant expansion of 7 \kms, we get
approximately 15 Myr to clear out a depression of that size.  

We then tried to estimate the amount of energy required to form such a
depression using the method in \citet{chevalier74}. Here we assume a
spherically symmetric expansion. If the expansion has already blown out
of the disk, this may not be the case (and would increase the estimated
age of the feature; see below). For a velocity of 7
\kms\ and a radius of 105 pc, we find the total energy required to be
$E_s=1.6n_0^{1.12} \times 10^{51}$ ergs. To get an estimate for the
density, $n_0$, we assumed a pre-existing column density of $1.5 \times
10^{21}$ \acm\ based on the average density in the region around the
small depression, and a scale height for the galaxy of 200 pc. The scale
height was taken to be approximately equal to the radius of the larger
\HI\ depression/cavity (see below). This gives us $n_0=1.0$ cm$^{-3}$,
and a resulting total energy of $1.6 \times 10^{51}$ ergs. This is
similar to the $10^{51}$ ergs from a typical Type II supernova, so it is
feasible in this sense.  We then attempted to apply the models of
\citet{mccray87} for the formation of supershells from stellar winds and
supernovae to estimate the number of stars with masses greater than
7\msun\ required to produce a hole of a given size. The resulting number
was approximately 200, of which about 50 should be O stars. However, the
resulting timescale for shell formation calculated from the models was
short, only 2 Myr. In 
this short time, the expansion due to the star formation is still
primarily from stellar winds; and it is unlikely that any stars would
yet have exploded as supernovae, and we should be able to see an OB
association of this size---which we don't. So about all we can conclude
is that the x-ray source appears to be coincident with an expanding
depression in the \HI; and adjacent to a region of disturbed gas to the
south.

We also examined the kinematics of the gas around the edges of the
larger depression/cavity in the \HI. However, no expansion was
detected. If we now use the average velocity dispersion in the gas (9
\kms) as an upper limit to any expansion, with a radius of 205 pc we
obtain a lower limit of 22 Myr for a hole this size to form. Because we
don't detect expansion, the hole could be stalled because it has broken
out and is freely expanding perpendicular to the disk, and so could be older
than that. We used the same methods to estimate the required energy and
number of stars required to form such a depression as for the area
around the x-ray source. This time, we used a column density of $2.5
\times 10^{21}$ \acm\ based on the existing gas density around the
cavity; giving us an $n_0$ of 1.6 cm$^{-3}$. The required energy is then
$3 \times 10^{52}$ ergs (equivalent to approximately 30 SNe). However,
from the \citet{mccray87} equations, the estimated number of stars
needed to create such a large depression via winds (and possibly
supernovae, but see the timescale discussion, below) with
masses greater than 7\msun\ is then 2800, which is extremely high. If
there were a young cluster that large, it would be easily visible. This
suggests an older age for the formation of the large \HI\ cavity; or
perhaps the gas was slowly depleted/blown-out over long time-scales by a
series of star-forming events rather than by one large cluster.

\begin{figure}
\epsscale{0.75}
\plotone{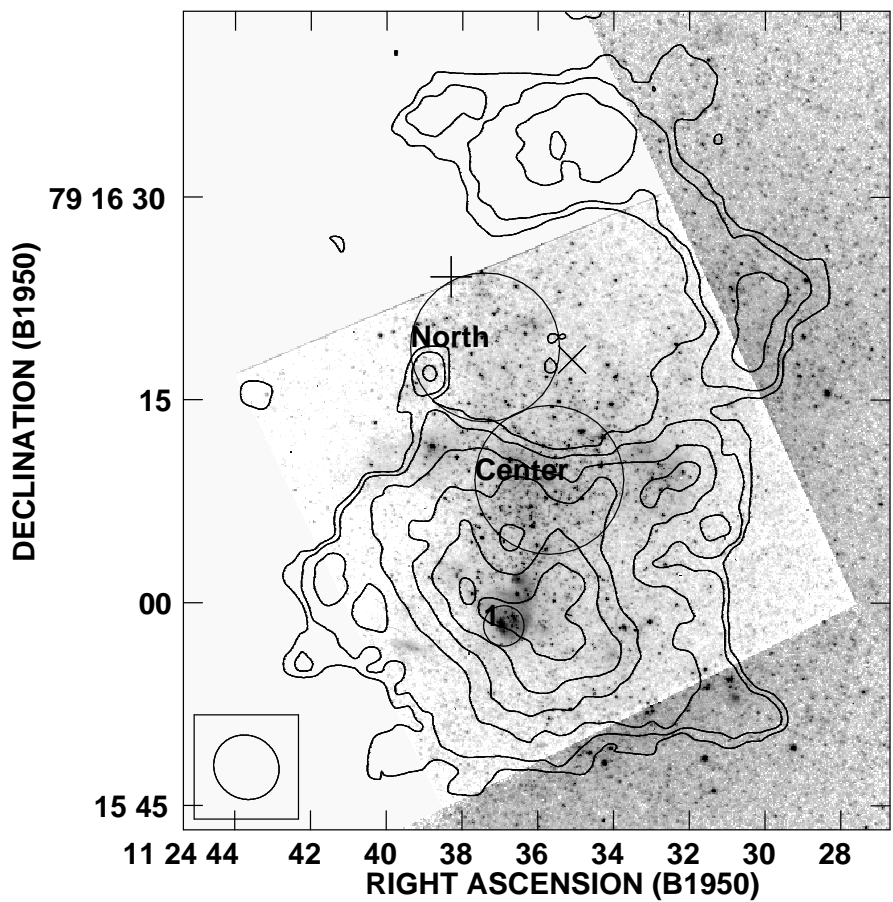}
\caption{The \protect\hres\ \protect\HI\ integrated flux contours on the
  {\it HST} WFPC2 image from \citet{lynds98}. The large circles show the areas
  for which they determined their north and central region CMDs. The
  small circle labelled ``1'' shows the location of the brightest \HII\
  region examined by \protect\citet{lynds98}. The
  plus sign indicates the location of the unresolved x-ray source
  detected by \citet{ott05a}; the $\times$  marks the location of the
  kinematic center as determined from the \protect\lres\ data. The beam
  size, shown in the lower left box, is $110 \times 100$ pc. Contours
 are at $(5, 10, 20, 30, 40, 50) \times 10^{20}$ \protect\acm. The
  lowest contour is at 0.7$\sigma$, the second at 1.5$\sigma$. (A
  single-channel 2\protect$\sigma$\ detection integrated over 3 channels
  is equivalent to a column density of 6.76 $\times 10^{20}$ \acm, which
  we take as indicative of the noise level in the integrated map.) 
\label{fig:bcd-1m0onhst}
}
\end{figure}

A more serious problem with applying the \citet{mccray87} model is that
doing so results in a calculated timescale of only 3 Myr; which is not
only  short in stellar evolution terms, but also inconsistent with
the lower limit of 22 Myr we calculated, above. A similar disconnect was
found by \citet{lynds98} for formation of the large \HII\ regions in
\vii. They used the Wide Field Planetary Camera 2 (WFPC2) on the {\it
Hubble Space Telescope} (HST) to produce color-magnitude diagrams (CMDs)
of several of the bright \HII\ regions in \vii.  The brightest \HII\
region (visible as the darkest region of the grayscale in
Figure~\ref{fig:bcd0m0onha} and marked in
Figure~\ref{fig:bcd-1m0onhst}) was consistent with an OB association
containing approximately 25 bright blue (O type) stars; the age of the
cluster based on the CMD was 4--5 Myr. From their ground-based long-slit
emission line spectra of the \HII\ regions, they were able to detect
expanding \ha\ shells; including a partial shell around the brightest
\HII\ region with a diameter of 79 pc and an expansion velocity of $\sim
65$ \kms. They then attempted to use the \citet{mccray87} models as we
did, but they also found a timescale that was too short: $\sim$ 1 Myr,
which did not match the 4--5 Myr age estimate from the CMD. They
concluded that perhaps this indicated that the starburst that formed the
association was not instantaneous \citep{shull95}.

\begin{figure}
\epsscale{0.85}
\plotone{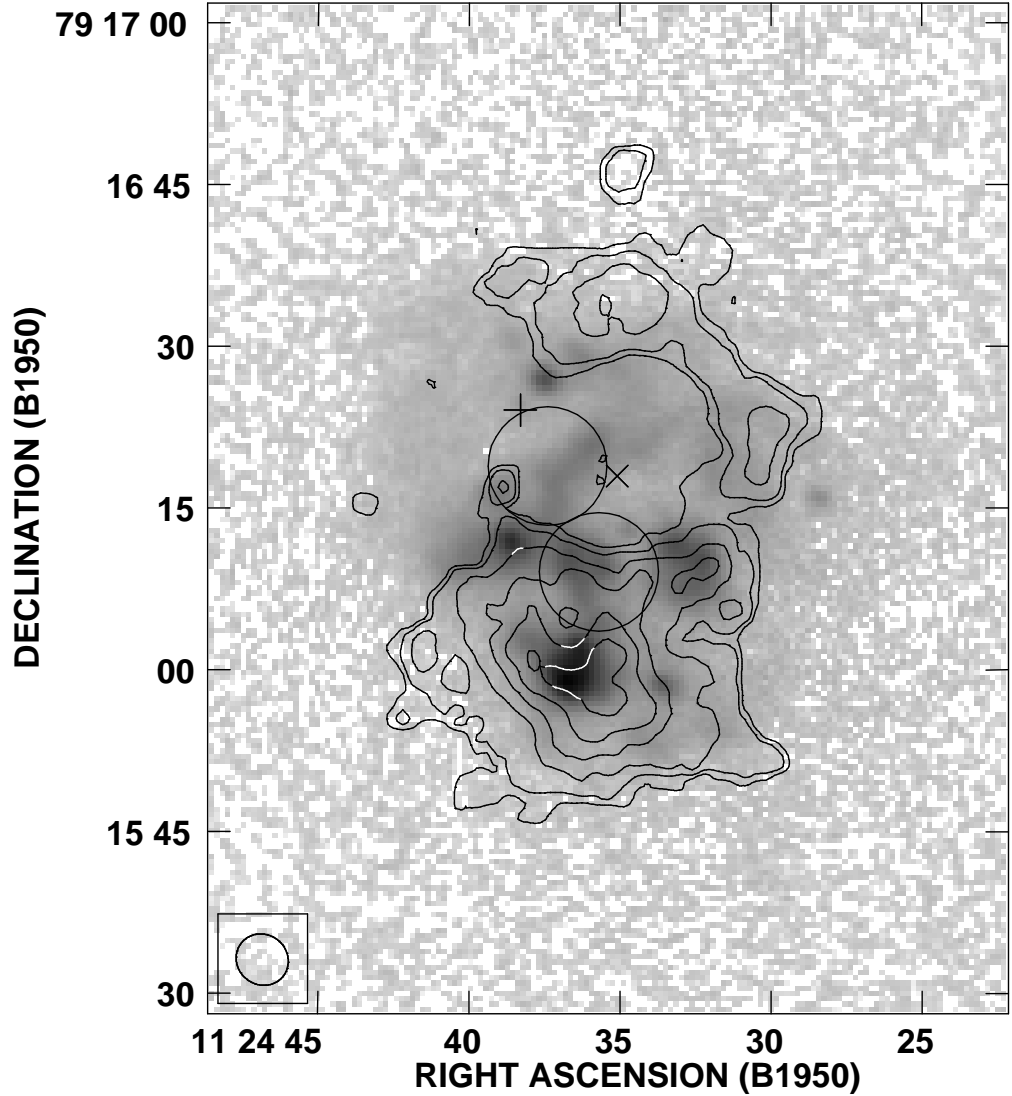}
\caption{The \protect\hres\ \protect\HI\ integrated flux contours on our
  \protect\ha\ image. The circles show the areas
  for which \protect\citet{lynds98} determined their north and central
  region CMDs. The 
  plus sign indicates the location of the unresolved x-ray source
  detected by \citet{ott05a}; the $\times$  marks the location of the
  kinematic center as determined from the \protect\lres\ data. The beam
  size, shown in the lower left box, is $110 \times 100$ pc. Contours
  are the same as for Figure~\protect\ref{fig:bcd-1m0onhst}. 
\label{fig:bcd-1m0onha}
}
\end{figure}

\citet{lynds98} also produced CMDs of some of the central areas of
\vii. The location of their ``center'' and ``north'' fields are marked
in Figure~\ref{fig:bcd-1m0onhst}, which shows the HST image with
contours from the \hres\ integrated flux map. Neither field corresponds
exactly with the \HI\ cavity/depression, but the ``north'' field is
located on the east side of it, and just south of the x-ray source. The
CMD for the ``north'' field is less rich in massive stars than the one
for the ``center,'' which revealed approximately 20 red supergiants ($>
10$ Myr) and some blue supergiants ($\sim 5$ Myr), but both regions
contain a mixed-age population of stars with ages of 5 and 10 Myr. In
general, \citet{lynds98} find a relatively smooth elliptical
distribution of the evolved stars with a stellar surface density that
falls off exponentially. The younger, bluer massive stars tend to
exhibit clustering into associations and groups, and are primarily
located in the central regions of the galaxy. The youngest are
associated with the bright \HII\ regions, which we find are co-located
with \HI\ peaks (Figure~\ref{fig:bcd-1m0onha}). They posit star
formation 1--2 Gyr ago followed by a strong starburst episode 600--800 Myr
ago, with less vigorous star-formation since then. Interestingly, the
kinematic center of \vii\ as identified by the rotation curve analysis
lies  close to the center of the large \HI\ cavity. If the cavity is
the result of star-formation, this would be consistent with
centrally-concentrated star-formation activity in the past. There is
evidence that stellar feedback from star formation events can and does
create sometimes large ($\approx$ 1 kpc) \HI\ holes and shells, as found
by \citet{weisz09} in IC 2574.

Another possibility for forming the hole is consumption of the gas by
star formation. The timescale to consume the gas that filled the hole at
the current star formation rate, given our estimate of the pre-existing
level of the gas in that region, is of order 700 Myr. This is
relatively short if star formation has been on-going in this region for
that extended period of time. Alternatively, the original gas in this
region could have been consumed quite quickly by the starburst that took
place 600--800 Myr ago \citep{lynds98}, but we then require that
the hole not be replenished very rapidly, which seems unreasonable given
that the hole should have filled in long ago just from the dispersion in
the gas.

\subsection{Origin of the Starburst}

One possible explanation for the enhanced star formation episode that
characterizes BCDs is proposed by \citet{salzer99}, who suggest that it
is a natural consequence of the central mass density concentration found in
BCDs. In an optical study of a sample of 18 BCDs and 11 dIrrs, they find
that when the luminosity contribution from the starburst is ignored (by
fitting the underlying host galaxy's surface brightness profile as
determined from an exponential fit to only the outer portions), the
underlying hosts of BCDs have shorter scale lengths and higher central
surface brightnesses than dIrrs with similar absolute magnitudes. They
cite the shorter scale lengths as evidence of stellar compactness: that
BCDs are low-surface brightness galaxies with the highest central
stellar mass concentrations in the dwarf spectrum.

There is some evidence that this may be the case for the \HI\ gas
in BCDs as well: \citet{vanzee98} plotted \HI\ surface density profiles
for a sample of BCDs and dIrrs, and found that the gas in BCDs is more
centrally concentrated than in similar dIrrs. It is commonly thought that
the gas surface density must exceed a critical threshold in order to
support a starburst\footnote{Although azimuthal averages of the gas
density in dwarfs rarely exceed this critical value, there is evidence
that local values in the region of active star formation do
\citep{hew01,vanderhulst93,vanzee97,meurer98,deblok06}.}
\citep{kennicutt89}; \citet{salzer99} also suggest that the high central
mass density concentration allows these galaxies to retain their gas, allowing
episodic bursts, and calculations by \citet{maclow99} indicate that 
mass loss should be minimal for galaxies with gas masses greater than
$10^7$ \msun. A follow-up study of the \HI\ in a sample of BCDs by
\citet{salzer02} found that the $M_{HI}/L_B$ for these galaxies is about
twice as high as that for low surface brightness dwarfs  when
only the luminosity of the underlying low-surface brightness component
of the BCD is taken into account. This provides a significant gas
reservoir to support the starburst episodes, and could enhance the
ability of the galaxy to retain gas during/after a burst. Additionally,
if the gas remains concentrated even after a burst, then the timescale
between bursts could be  short. This would make it difficult to find
a BCD in the ``off'' state.

If this hypothesis is correct then we would expect \vii\ to exhibit
signs of central compactness, both optically and in the \HI. We find
that compared to the sample in \citet{salzer99}, \vii\ has a slightly
lower $\mu_{B_0}$ at a given M$_B$ than their sample of BCDs, although
slightly higher than for their LSBDs. As mentioned in
\S\ref{sec:optical}, \vii's scale length (relative to its $M_V$) lies at
the peak of the distribution of the sample of dIms and BCDs plotted in
\citet{he06}. Compared to the \citet{salzer99} sample, however, it is
large not only for their sample of BCDs, but also compared to their LSBD
sample. So the underlying stellar component does not appear to be
greatly centrally concentrated. However, there are some indications for
central concentration in the \HI\ on larger scales (i.e.\ at a linear
resolution similar to that for which structure is seen in the \HI\ in
dIrrs; see Figure~\ref{fig:cd1m0onv}).

However, since the vigorous star formation episodes that characterize
BCDs are relatively short-lived compared to the age of the host galaxy,
perhaps these systems don't start out with a high central mass
concentration. \Citet{vanzee01} finds that BCDs are low angular momentum
systems, and suggests that they have collapsed to more centrally
concentrated systems with steeper rotation curves. This affects the
critical threshold density for star formation, and makes BCDs more
susceptible to star bursts than higher angular momentum systems. This
``collapse'' requires a redistribution of mass in the galaxy.  One proposed
mechanism for mass redistribution and angular momentum transfer in low
mass BCDs (without nearby companions) is a barlike torque arising from
the irregular symmetry and irregular mass distribution in these small
systems, or from an elongated dark matter halo \citep{he04}. Such mass flow
should show up in the rotation pattern of the gas disk as a central
twisting of the isovelocity contours. \vii\ has a change of
15--20\arcdeg\ between the NE and SW parts of the galaxy, so we can't
rule out the possibility of some kind of mass flow in the gas disk.

Another possible mechanism for flattening the gravitational potential
and expansion of the stellar component is suggested by \citet{papa96b},
who propose that the heating of the ISM by the energy from the starburst
could be the mechanism for mass and angular momentum
redistribution. Although \vii's scale length is not particularly short,
which could indicate that it has undergone some expansion, the lack of
diffuse x-ray as well as the lack of extensive large-scale structure in
the \HI\ (the \HI\ depression discussed above occupies only a small,
central part of the galaxy) would seem to indicate that not much energy
has been injected into the galaxy as a whole by star formation. It seems
unlikely, then, that enough energy would have been produced to initiate
any significant redistribution of the mass in the galaxy, especially the
stellar component. In conclusion, it isn't clear that this theory of
large-scale mass re-arrangement can explain what we see in \vii.

Interactions, either with other galaxies or with small gas companions,
have also been proposed as the triggering mechanism for the starbursts
that occur in BCDs (see \citealt{brosch04} for an overview of the
literature), yet surveys to detect nearby possible perturbers around
BCDs have had mixed results
\citep[e.g.][]{campos91,telles95,taylor97,putman98,telles00,pustilnik01,pustilnik03,brosch04,he04},
\vii\ would seem to be in the set of galaxies that haven't been tidally
triggered, as we have found no nearby \HI\ clouds or companion
galaxies. However, if it had accreted a small perturber in the distant
past, this might be difficult to detect.

There is observational evidence that some dwarf systems have undergone
accretion or merger events in the past. A nearby example is thought to
be IC 10, which \cite{wilcots98} argue still has primordial gas settling
into the galaxy via accretion. Classic signatures of accretion include
the appearance of two distinctly different kinematic components, visible
in the \HI\ or stellar velocity fields. Although the gas isovelocity
contours for VII Zw 403 are somewhat perturbed, this isn't unusual for
BCDs (\S \ref{sec:kinematics}), and there don't appear to be two
completely separate components despite the change in position angle of
the \HI\ kinematic axis. Another accretion signature is a misalignment
between the optical and \HI\ axes. \citet{hunter02} plotted the
difference between optical and \HI\ kinematic axes for 47 Im and Sm
galaxies taken from \citet{swaters99} and found that about half had
offsets less than 10\arcdeg, with the maximum being 40\arcdeg. For VII
Zw 403, the optical axis is at $-11$\arcdeg. In the NE half of the
galaxy, the \HI\ kinematical axis is at +45\arcdeg\ to +50\arcdeg, and is around
+20\arcdeg\ to +30\arcdeg\ in the SW part, so there is a somewhat significant
misalignment between the gas and optical components of between
30\arcdeg\ and 60\arcdeg.

For further comparison, we look to the gas-rich Im galaxy DDO 26, for
which \citet{hunter02} conclude that the most likely explanation for the
lack of relaxation they detect in its gas kinematics is the on-going
merger of two small gas-rich objects. They find an optical/\HI\ axis
misalignment of 60\arcdeg, along with a change in the \HI\ position
angle with radius. They also find that the velocity field of DDO 26 is
fairly regular in the inner regions and becomes less so in the outer
disk. These are features that are similar to but more extreme than those
we see in \vii. Differences include the presence of a faint \HI\ arm in
DDO 26 and the detection of two separate peaks in the \HI\ intensity
which were both fairly broad ($\sim 40$ \kms) in velocity. Additionally,
DDO 26 is currently quiescent, with a SFR of 0.0004 \msun\
yr$^{-1}$ kpc$^{-2}$ compared to \vii's rate of 0.02 \msun\ yr$^{-1}$
kpc$^{-2}$. So from these comparisons, it seems possible that \vii\
underwent an accretion or merger event in the past.

Another possibility to explain the skew in the \HI\ kinematic axis is
that expansion of the \HII\ regions has caused a subsequent expansion in
the surrounding \HI, lifting it out of the plane of the galaxy. If the
near side of the galaxy is to the west (right), and we envision the
expanding \HI\ as distributed in a torus aligned with the plane of the
galaxy, then gas to the left of the kinematic center would be
blue-shifted relative to the bulk of the galaxy. Likewise, the gas to
the right of center would be somewhat behind the bulk of the galaxy and
red-shifted. The expanding gas would then skew the observed velocity
field in the sense that we see: the isovelocity contours in the north
part will be skewed to the left, and those in the south skewed to the
right.  If the expansion is on the order of the 7 \kms\ we measured for
the lowest density region around the x-ray source, or the 9 \kms\ limit
set by the average dispersion, then in this slowly rotating galaxy, the
expected skew would be large, which is what we see. In this case, the
hole in the \HI\ would be due to a combination of consumption and slow
blow-out.

\section{Summary}

We have used multi-wavelength data to examine the relation between the
stars and gas in \vii\ in an effort to understand the origin of the
starbursting activity that characterizes BCD galaxies. As for many BCDs,
in the optical $V$-band we find a mostly symmetrical, elliptical
distribution of stars; this is overlaid with a color gradient that is
bluer in the central regions and redder in the outer. This indicates
that the central regions have undergone star formation in the past 200
Myr, whereas the outer region colors indicate constant star formation
over the past 10--20 Gyr. The galaxy also has a few large \HII\ regions
concentrated toward the center of the galaxy (within the inner 1.3 kpc)
where star formation has occurred in the past 10 Myr. Most of the
activity is in a large complex just south of the central part of the
galaxy. Overall, the SFR for \vii\ is at the high end of the normal
range for dIrrs, which is, again, typical for BCDs. However, the galaxy
can continue to form stars at this rate for at least 5 Gyr more; 10 Gyr
if we include gas recycling \citep{brinchmann04}.

Our new VLA data in combination with earlier data resulted in a lower
total \HI\ flux detection than single-dish estimates; we believe this is
due to inclusion of Milky Way emission in the single-dish spectra and in
the low-resolution VLA observations. We have attempted to remove this
Galactic contamination from our low-resolution (D configuration) data
for our analysis here. Our total \HI\ mass estimate then is $5 \times
10^7$ \msun, which is lower than the median for the Salzer BCD sample
\citep{salzer02}, but within the normal range of values. However, the
\mhilb\ indicates that \vii\ is somewhat gas-rich for a BCD of similar
luminosity. Conversely, it may be ``luminosity-poor'' relative to other
similarly-sized BCDs: earlier work by \citet{lynds98} concluded that the
SFR for \vii\ was up to 30 times greater 600--800 Myr ago. Perhaps we
are seeing it post-burst. The idea that \vii\ is presently past the peak
of a starburst event 
could also explain the non-detection of diffuse x-ray emission by
\citet{ott05a,ott05b}, indicating a lack of coronal wind or that it has
had time to cool.

The \HI\ velocity field, although complex with a change in position
angle of the kinematic axis between the NE and SW components of about
15--20\arcdeg, is not that unusual for a BCD. We see no other strong
indications of a large-scale gravitational interaction in the velocity
field, such as tidal tails or strewn-about \HI. The skew in the velocity
field can, however, be explained if the \HI\ in the galaxy is undergoing
a general symmetric expansion away from the plane on both sides
surrounding the star-forming region as a result of pressure from the
\ha\ region. The rotation curve is essentially solid-body for both
halves of the galaxy, and with correction for asymmetric drift, produces
a dynamical mass estimate that is similar to those of other BCDs
\citep{taylor94,vanzee98,vanzee01}.

If the galaxy is post-starburst, then one might expect to see
some effect in the interstellar medium from the energy injected by
stellar winds and supernovae. Additionally, star formation should use up
some of the neutral gas in the galaxy. In accordance with this, although
the \HI\ is more centrally concentrated in \vii\ than in dIrrs, the
central \HI\ surface density is lower than is typical for the BCDs in
the \citet{vanzee98} sample, and the \HI\ is more extended---out to 4.3
R$_{25}$ (2.3 R$_H$). At large scales (650 pc), the \HI\ distribution
appears relatively smooth as for most BCDs.  This is in contrast to
dIrr galaxies, which at the same scale often show structure in the
gas component in the form of holes, shells, and ridges. For \vii\
however, we begin to see some structure in the \HI\ distribution only at
small (200 pc) scales. 

There is clearly a connection between recent star formation and the \HI\
distribution and kinematics in the galaxy. The youngest stars (4--5 Myr
old) are located near \HII\ regions and local peaks in the \HI, with the
regions of highest \HI\ column density corresponding to the regions of
strongest \ha\ flux (as seen in other BCDs). Our \HI\ maps also show
that the regions of higher dispersion are located near, although not
completely coincident with, with \ha\
regions; perhaps in response to the energy injected by the star
formation activity.  Interestingly, the region of highest dispersion in
our map is directly adjacent to the unresolved x-ray source detected by
\citet{ott05a} that could be an x-ray binary. The source is located near
an \HII\ region, and is in a small lower-density region at the edge of a
larger depression or cavity in the \HI. The center of the large \HI\
cavity is coincident with the kinematic center of the \HI\ velocity
field.

We attempted to investigate whether the small low-density region could
be the result of star formation/SNe blowing a hole in the gas.  Although
both the amount of energy required and the number of massive stars
needed were reasonable, the timescale from the shell formation model was
too short for many SNe to have already occurred. We performed the same
analysis for the (stalled) large cavity in the \HI, but found that in
addition to requiring the presence of a large star cluster (not seen),
the timescale from the model was again too short. Similar timescale
discrepancies were found by \citet{lynds98}, leading them to invoke
non-instantaneous starburst/star-formation activity in a series of
events.

So we are brought back to the question of what triggered the vigorous
star formation episode that apparently peaked in the past but is still
ongoing today. We have investigated three possible triggers: tidal
forces from an external perturber, inherent central mass density concentration,
and accretion of a smaller object. We found no nearby large or small
perturbers, so it has apparently not been tidally triggered. \vii's mass
doesn't appear to be strongly centrally concentrated either, as
evidenced by its optical scale length, lower central surface
brightness, and lower \HI\ surface density than comparable BCDs. This
leaves us with an accretion event.  There is evidence for accretion in
some dwarf systems, including some Local Group dwarfs. Compared to
another suspected accretion/merger system (DDO 26), \vii\ has some
similarities but less extreme signatures. Of the three possible triggers
we have investigated, this seems the most likely; although the evidence
is far from conclusive.


\acknowledgments

This publication makes
use of data products from the Two Micron All Sky Survey, which is a
joint project of the University of Massachusetts and the Infrared
Processing and Analysis Center/California Institute of Technology,
funded by the National Aeronautics and Space Administration and the
National Science Foundation.  This work is also based in part on
archival data obtained with the Spitzer Space Telescope, which is
operated by the Jet Propulsion Laboratory, California Institute of
Technology under a contract with NASA.

We appreciate the use of Ohio State University's Fabry-Perot and the
Ohio State Infrared Imager-Spectrograph and Mark Wagner for making
them work for us. The authors would like to thank the anonymous
referee, who provided valuable comments and criticism that improved the
paper. CES would like to thank Glenn Morrison and Frazer
Owen for help with data calibration questions.

Support for CES came from grants AST-0407051 and AST-0707468 from the
National Science Foundation. Support to DAH and TEN for this research
came from the Lowell Research Fund and grants AST-9802193 and
AST-0204922 from the National Science Foundation. Support to DAH also
came from NASA through grant NNX07AJ36G. Funding to BGE was provided
by NSF grant AST-0707426. Support for TA came from the National
Science Foundation Research Experience for Undergraduates program
(AST-0552798) in partnership with the Department of Defense ASSURE
programs (Awards to Stimulate and Support Undergraduate Research
Experiences).

Facilities: \facility{VLA} \facility{GALEX} \facility{Lowell Observatory}

\clearpage
 
\begin{deluxetable}{lccc}
\tablecaption{VLA Observations
\label{tab:HIObs}}
\tabletypesize{\small}
\tablewidth{0pt}
\tablehead{
\colhead{Parameter}	& \colhead{D configuration}	
				& \colhead{C configuration}
                                        & \colhead{B configuration}    
}
\startdata

Observation Date 	& 1997 Nov.\ 10	   & 1992 April 11 & 2006 Sept.\ 10 \\  
Time on Source~(min) 	& 253 		   & 220 	   & 546            \\
Central Velocity (\kms)	& $-60$		   & $-60$       & $-60$\tablenotemark{a} \\
Bandwidth~(MHz) 	& 1.56 		   & 1.56 	   & 1.56	    \\
No.\ of Channels	& 128 		   & 128 	   & 128	    \\
Velocity Resolution (\kms)& 2.6 	   & 2.6 	   & 2.6	    \\
Beam Size\ (arcsec) 
                        & $72.91 \times 53.57$ 
                                           &\nodata
                                                           &\nodata       \\
Single Channel r.m.s.~(mJy beam$^{-1}$)
                        & 0.93 		   & \nodata	   & \nodata     \\
Single Channel r.m.s.~(K)& 0.28 	   & \nodata 	   & \nodata \\
\enddata
\tablenotetext{b}{Doppler tracking not available; central fixed frequency
  chosen to correspond to $-60$ \kms. See text for details.}
\tablenotetext{a}{Natural weighting}
\end{deluxetable}

\clearpage

\begin{deluxetable}{lcccc}
\tablecaption{Combined Configuration Parameters
\label{tab:BCDdata}}
\tabletypesize{\small}
\tablewidth{0pt}
\tablehead{
\colhead{Parameter}	& \colhead{\protect\lres}
                                & \colhead{\protect\bcdone}
                                                       &\colhead{\protect\mres} 
                                                                   & \colhead{\protect\hres}}
\startdata

Robustness Factor
                        & +1		   & +1		   & 0                & $-1$\\
Beam Size\ (arcsec) 
                        & $34.10 \times 27.60$ 
                                           &$12.85 \times 11.71$ 
                                                           &$9.23 \times 7.57$ 
                                                                        &$ 4.95 \times 4.63$ \\
Beam Size\ (pc) 
                        & $730 \times 590$ 
                                           &$280 \times 250$ 
                                                           &$200 \times 210$ 
                                                                        &$ 110 \times 990$ \\
Single Channel r.m.s.~(mJy beam$^{-1}$)
                        & 0.83 		   & 0.80 	   & 0.86
& 0.90 \\
\enddata
\end{deluxetable}

\clearpage
 
\begin{deluxetable}{lc}
\tablecaption{Summary of Properties. 
\label{tab:global}}
\tablewidth{0pt}
\tablehead{
\colhead{Parameter} & \colhead{Value} 
}
\startdata
R.A. (B1950; $V$) \dotfill & $11^{\rm h}\ 24^{\rm m}\ 33\fs4$ $\pm$4\arcsec \\
Decl. (B1950; $V$) \dotfill & 79\arcdeg\ 16\arcmin\ 10\arcsec$\pm$ 3\arcsec \\
R.A. (B1950; \HI) \dotfill & $11^{\rm h}\ 24^{\rm m}\ 35\fs1$$\pm$ 10\fs3 \\
Decl. (B1950; \HI) \dotfill & 79\arcdeg\ 16\arcmin\ 18\arcsec $\pm$ 27\arcsec \\
$b/a$ ($V$) \dotfill & 0.49 \\
P.A. ($V$) \dotfill & $-11$\arcdeg \\
$D$ (Mpc) \dotfill & 4.4 \\
$M_{\rm HI}$ (M$_\sun$) \dotfill & $5.23\times 10^7$ \\
$R_{\rm HI}$\tablenotemark{a} (arcsec, kpc) \dotfill & 157, 3.35 \\
$E(B-V)_f$\tablenotemark{b} \dotfill &  0.023 \\
$R_{25}$ (arcsec, kpc) \dotfill &  36.5, 0.78 \\
$R_H$ (arcsec, kpc) \dotfill &  68.95, 1.47 \\
$\mu_0^V$ (magnitudes arcsec$^{-2}$) \dotfill & 23.12$\pm$0.09 \\
$R_D^V$ (kpc) \dotfill &  0.50$\pm$0.01 \\
$M_{V_0}$ (magnitudes)  \dotfill & $-$14.26$\pm$0.009 \\
($U-B$)$_0$  \dotfill & $-$0.50$\pm$0.01 \\
($B-V$)$_0$  \dotfill & 0.27$\pm$0.01 \\
($V-J$)$_0$  \dotfill & 1.30$\pm$0.01 \\
(FUV$-$NUV)$_0$ \dotfill & 0.17$\pm$0.01 \\
(NUV$-$V)$_0$\tablenotemark{c}~ \dotfill & 0.70$\pm$0.01 \\
log $L_{H\alpha,0}$ (ergs s$^{-1}$) \dotfill & 39.34$\pm$0.001 \\
log $L_{NUV,0}$ (ergs s$^{-1}$ Hz$^{-1}$) \dotfill & 26.07$\pm$0.001 \\
SFR$_{H\alpha}$\tablenotemark{d}~ (M\protect\solar yr$^{-1}$) \dotfill & 0.015 \\
SFR$_{FUV}$\tablenotemark{d}~ (M\protect\solar yr$^{-1}$) \dotfill & 0.013 \\
SFR$_{\rm D, H\alpha}$\tablenotemark{e}~(M\protect\solar yr$^{-1}$ kpc$^{-2}$) \dotfill & $-$1.72 \\
\enddata
\tablenotetext{a}{Measured to $N_{\rm HI} = 1 \times 10^{19}$ \protect\acm.}\\
\tablenotetext{b}{E($B-V$)$_f$ is foreground reddening due to the Milky Way
(Burstein \& Heiles 1984).
For the stars in VII Zw 403, we assume an additional internal reddening of 0.05
magnitude; for the \protect\HII\ regions
we assume an additional internal reddening of 0.1 magnitude.}
\tablenotetext{c}{$NUV$ is an AB magnitude.}
\tablenotetext{d}{Star formation rates derived from L$_{H\alpha}$ or the ultraviolet
using the formulae of \protect\citet{hunter10} that integrates
from 0.1 M\protect\solar to 100 M\protect\solar with a
\protect\citet{salpeter55} stellar initial mass function.}
\tablenotetext{e}{Star formation rate normalized to the area within
one scale-length $R_D^V$.}
\end{deluxetable}

\clearpage

\clearpage

\clearpage

\clearpage

\clearpage

\clearpage

\clearpage

\clearpage

\clearpage

\clearpage

\clearpage

\clearpage

\clearpage

\clearpage

\clearpage

\clearpage

\clearpage

\clearpage

\clearpage

\clearpage

\end{document}